\documentclass[twocolumn]{aastex631}
\usepackage{graphicx}	
\usepackage{amsmath}	
\usepackage{booktabs}
\usepackage{xtab}
\usepackage{multirow}
\usepackage[left,displaymath,mathlines]{lineno}

\shorttitle{MONK simulations for different coronal geometries}
\shortauthors{Fan et al.}

\begin{document}

\title{Possible Coronal Geometry in the Hard and Soft State of Black Hole X-ray Binaries \\ from MONK Simulations}

\author[0009-0009-2549-1161]{Ningyue~Fan}
\affiliation{Center for Astronomy and Astrophysics, Center for Field Theory and Particle Physics, and Department of Physics,\\
Fudan University, Shanghai 200438, China}
\affiliation{Department of Physics, Stanford University, Stanford, CA 94305, USA}
\affiliation{Kavli Institute for Particle Astrophysics and Cosmology, Stanford University, Stanford, CA 94305, USA}

\author[0000-0002-3180-9502]{Cosimo~Bambi}
\affiliation{Center for Astronomy and Astrophysics, Center for Field Theory and Particle Physics, and Department of Physics,\\
Fudan University, Shanghai 200438, China}
\affiliation{School of Natural Sciences and Humanities, New Uzbekistan University, Tashkent 100007, Uzbekistan}

\author[0000-0002-5872-6061]{James~F.~Steiner}
\affiliation{Center for Astrophysics \textbar\ Harvard \& Smithsonian, Cambridge, MA 02138, USA}

\author[0000-0003-1702-4917]{Wenda Zhang}
\affiliation{National Astronomical Observatories, Chinese Academy of Sciences, Beijing 100101, China}

\correspondingauthor{Cosimo Bambi}
\email{bambi@fudan.edu.cn}

\begin{abstract}
Understanding the coronal geometry in different states of black hole X-ray binaries is important for more accurate modeling of the system. However, it is difficult to distinguish different geometries by fitting the observed Comptonization spectra. In this work, we use the Monte Carlo ray-tracing code \texttt{MONK} to simulate the spectra for three simple corona toy models widely proposed in observational studies: sandwich, spherical, and lamppost, varying their optical depth and size (height). By fitting the simulated \textit{NuSTAR} observations with the \texttt{simplcut*kerrbb} model, we infer the possible parameter space for the hard state and soft state of different coronal geometries. The influence of the disk inclination angle, black hole spin and coronal temperature is discussed. We find that in the lamppost model, if we exclude the case of a very extended corona, the disk emission is always dominant, making the lamppost geometry incompatible with the hard state. While the sandwich and spherical models can produce similar spectra in both the hard and soft states, the simulated \textit{IXPE} polarimetric spectra show the potential to break this degeneracy. Geometrical effects arising from the limited size of the corona become evident in lower-spin black holes and affect the spectral fitting, where the larger ISCO reduces the corona coverage of the inner disk.

\end{abstract}




\section{Introduction}

Black hole X-ray binaries (BHXRBs) are binary systems in which a stellar-mass black hole accretes mass from a companion star. During the accretion process, part of the gravitational energy of the accreted material is converted into electromagnetic radiation, primarily in the X-ray band for stellar-mass black holes \citep{Remillard_McClinktock2006ARA}. The X-ray radiation typically consists of a disk multi-temperature black body component, a corona Comptonization, and a disk reflection component \citep{bambi2018}.

The corona is usually described as an optically-thin thermal plasma around the black hole and the inner accretion disk, and the Comptonization spectrum is normally approximated by a power-law with a high-energy cut-off in the X-ray band \citep{Belloni2011}. As shown by analytical studies and Monte Carlo simulations, the slope of the power-law function and the cut-off energy depend on the coronal electron temperature and Compton optical depth \citep{Sunyaev1980,Zycki_Done_1999,Schnittman2010}. Therefore, these properties of the corona can be inferred by fitting the coronal spectrum with a cut-off power-law model, for example, the additive model \texttt{cutoffpl}. More physical corona models like \texttt{thcomp} \citep{Zdziarski2020_thcomp} and \texttt{simplcut} \citep{Steiner2017_simplcutx} directly include physical parameters such as corona covering (scattering) fraction or optical depth and temperature, in addition to the power-law index. These convolutional models provide a more consistent picture, because they link the disk seed photons and the Comptonized photons. Observational studies reveal that the typical value of the power-law index (photon index $\Gamma$) is 1.5--2.0 in the hard state and $\geq2$ in the soft state, and the high-energy cut-off is $\sim$ tens of keV \citep{Done2007, Remillard_McClinktock2006ARA}.

However, the impact of the coronal geometry on the Comptonization spectrum is more complex and is not very clear yet \citep{Schnittman2010}. \texttt{thcomp} and \texttt{simplcut} do not assume a certain coronal geometry when Comptonizing the disk seed photons. Therefore, we cannot directly constrain the coronal geometry from observations using these models. Meanwhile, oversimplifications of the coronal geometry may lead to inconsistent results in observational studies and more complex coronal geometries are sometimes proposed \citep[see, for example, ][]{Zdziarski2021_maxij1820,Kawamura2022_maxij1820}.

In this paper, we want to consider some specific coronal geometries and calculate the resulting X-ray spectrum of the BHXRB. The public code \texttt{MONK} \citep{Zhang2019_MONK} provides self-consistent Monte Carlo simulations of the photons emitted by a disk and scattered by a corona, incorporating different coronal geometries. It assumes a Novikov-Thorne emissivity profile for the disk \citep{Novikov1973} with zero torque at the inner edge of the disk and a color-corrected blackbody local spectrum. When the photons enter the corona, for each step the code evaluates the scattering probability assuming the Klein-Nishina \citep[see the original definition in][]{1929ZPhy...52..853K} or Thompson cross section. If a photon is scattered, the photon 4-momentum is recalculated. Finally, the code calculates the energy spectrum at infinity for different inclination angles of the disk with respect to the distant observer. 

Furthermore, the \texttt{MONK} code is also able to calculate the polarization spectrum of photons, including the polarization degree (PD) and polarization angle (PA). Simulating the polarization patterns of different coronal geometries can provide us with new insights into this problem, because the polarization spectrum is highly sensitive to the corona optical depth and geometry \citep{Beheshtipour2017,Schnittman2010,Zhang2022_lamppost_polarization,Krawczynski2022_kerrC}. Meanwhile, the X-ray polarization telescope \textit{IXPE} \citep{IXPE2016} was lunched in 2021, enabling us to compare the simulated polarimetric spectra with observations and constrain the coronal geometry. There are already some studies finding that \textit{IXPE} observations cannot be explained by simple disk or corona models, e.g., \cite{Krawczynski2024_4U1630,Ratheesh2024,Riaz2020_disklikecorona}. With the launch of the enhanced X-ray Timing and Polarimetry mission \citep[eXTP,][]{eXTP} in the future, we will have higher-quality polarization data to study this problem.

In this study, we use the \texttt{MONK} code to simulate different coronal geometries. By fitting the simulated energy spectra and studying the trend of the corona parameters, we try to constrain the possible coronal geometries for the hard and soft spectral state. In Section~\ref{method}, we introduce the simulation settings and fitting method. In Section~\ref{result}, we present the spectra of different coronal geometries and their fitting results. Then in Section~\ref{discussion}, we discuss the influence of the inclination angle, spin, and coronal temperature on the spectral parameters, and the potential of \textit{IXPE} to solve the degeneracy in energy spectra. The conclusions are in Section~\ref{conclusion}.

\section{Method}\label{method}

\begin{figure*}
    \centering
    \includegraphics[width=0.45\linewidth]{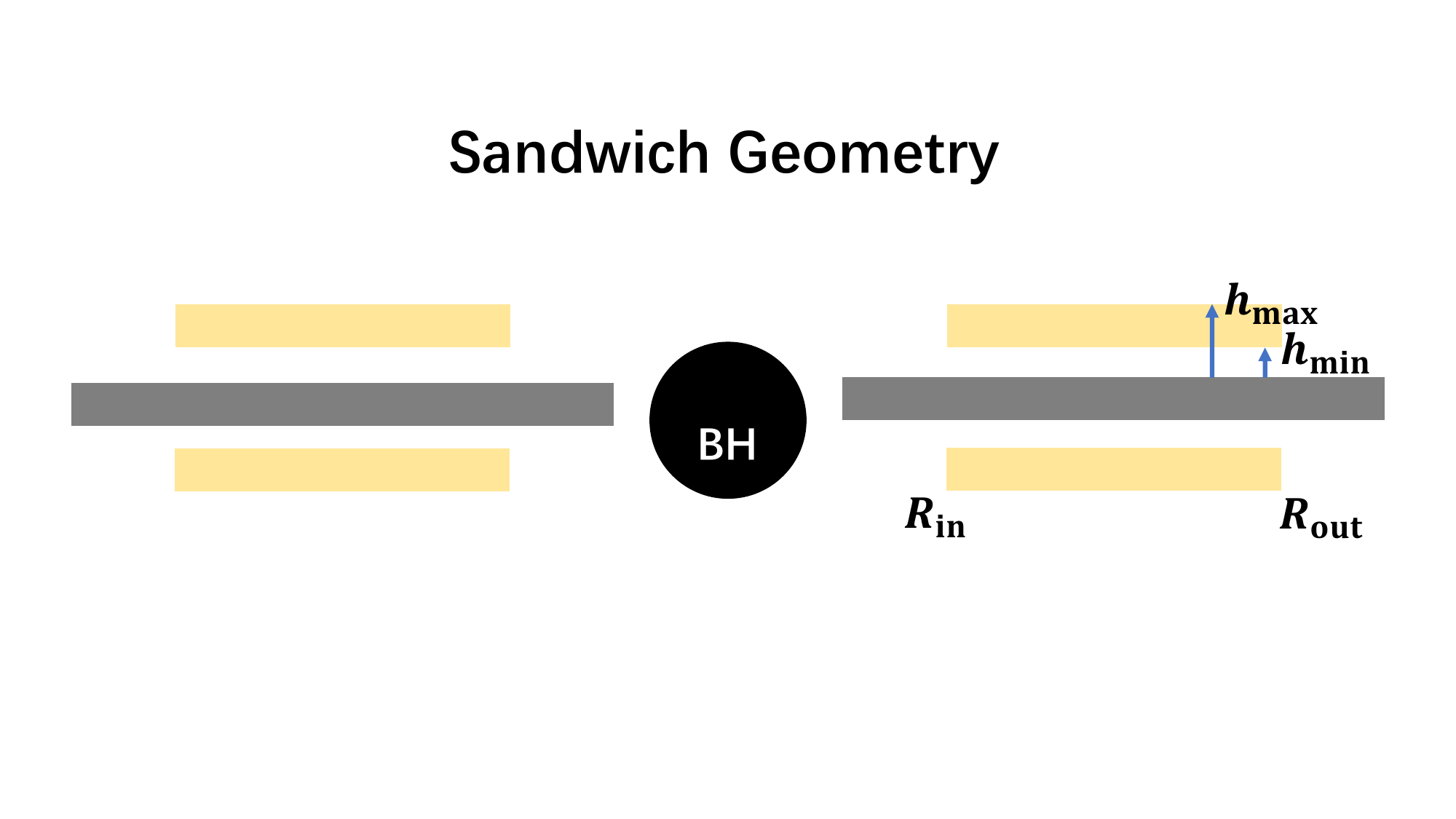}
    \includegraphics[width=0.45\linewidth]{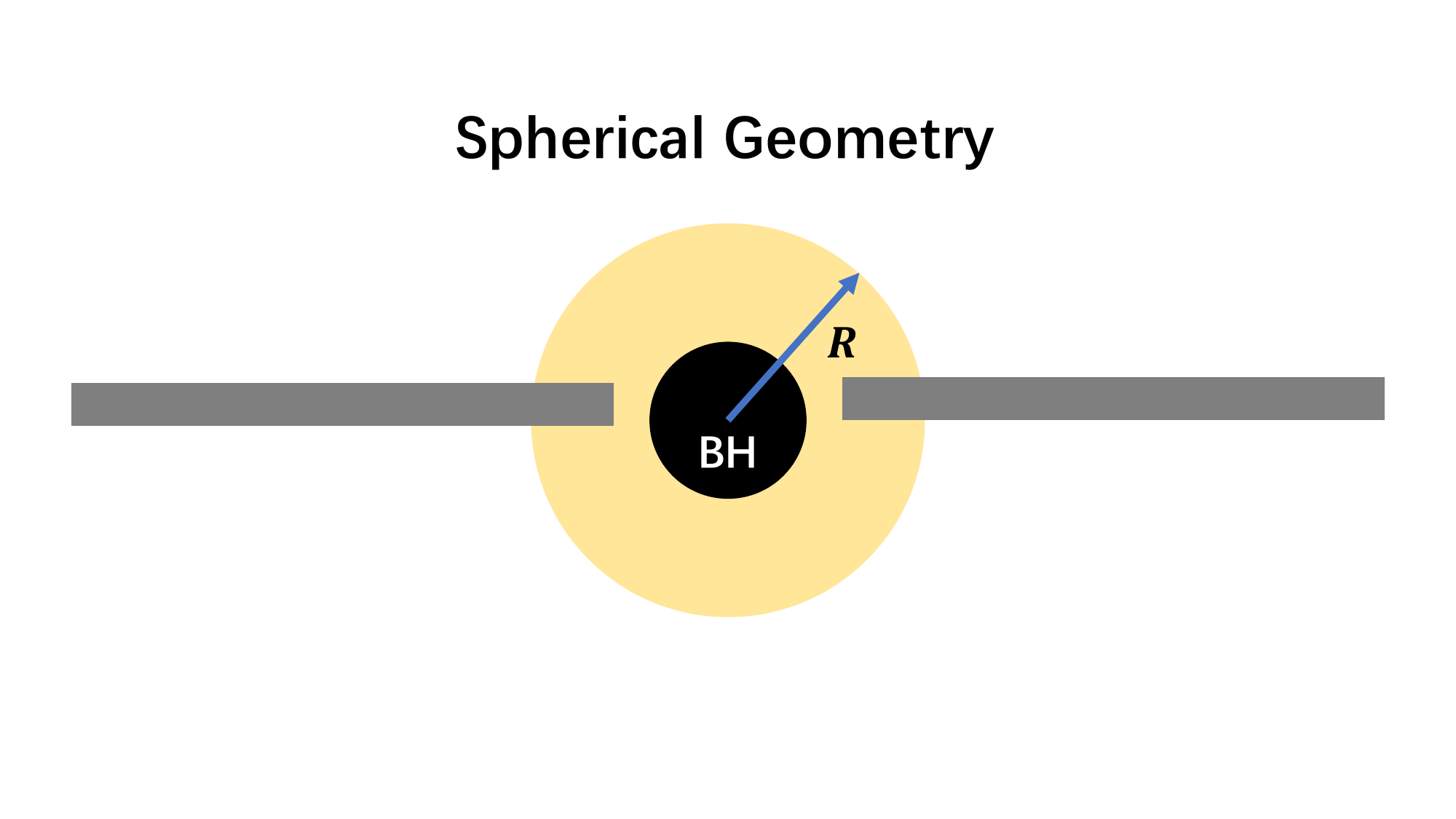}
    \includegraphics[width=0.45\linewidth]{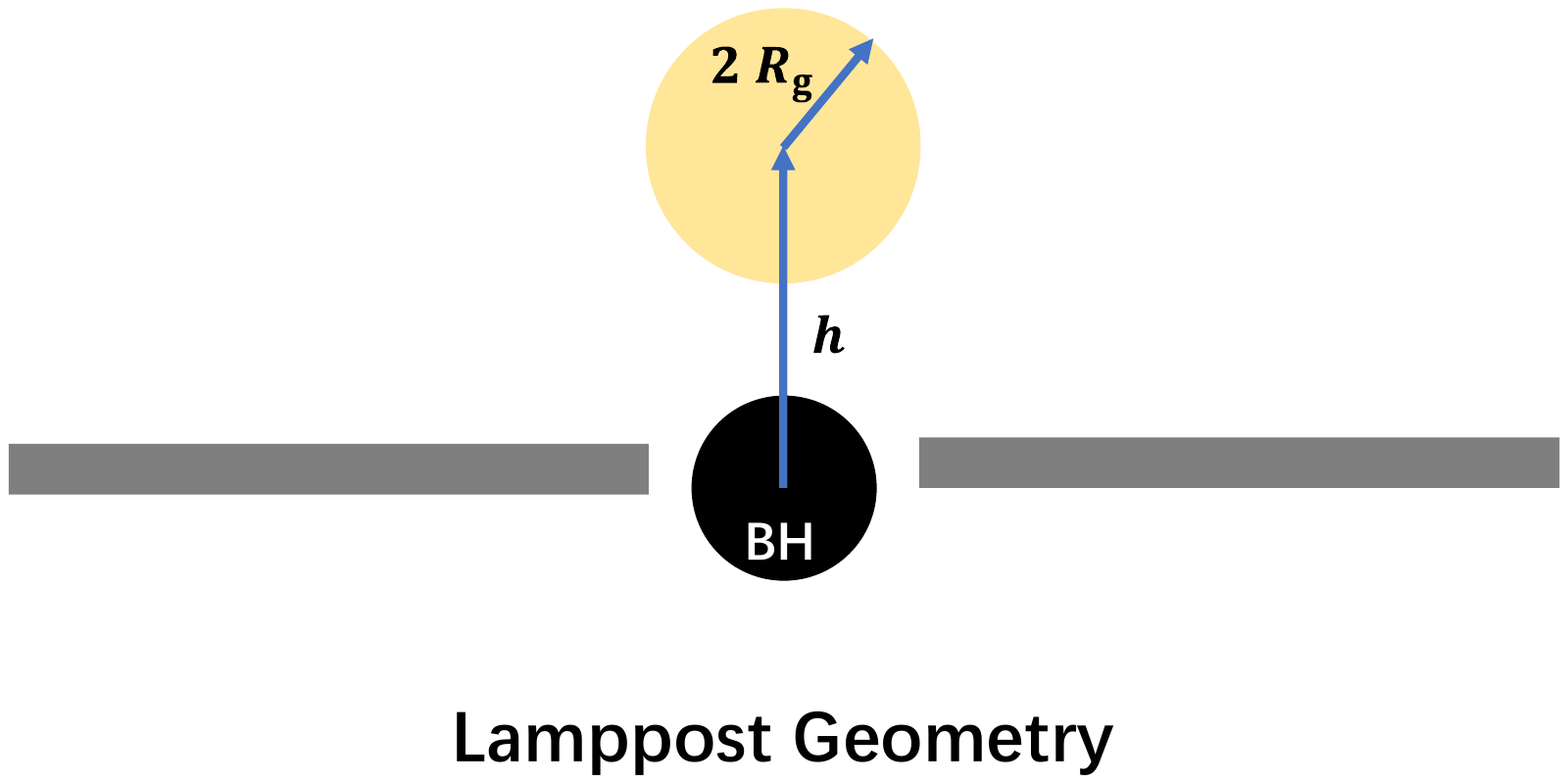}
    \caption{The coronal geometries considered in our study. The upper left panel is the sandwich corona, which is two parallel layers with a certain thickness above and below the disk. The upper right panel is the spherical corona, which is a central sphere around the black hole. The lower panel is the lamppost corona, which is a sphere of finite radius ($=2~R_g$ in our simulations) along the black hole spin axis.}
    
    \label{geometry}
\end{figure*}

Three types of coronal geometries are considered in this work: sandwich geometry, spherical geometry and lamppost geometry (see the illustration in Fig.~\ref{geometry}). These are widely proposed coronal geometries in theoretical and observational studies \citep{Bambi2017_geometry,Lohfink2017_geometry}. We change the geometry, size and optical depth of the corona while fixing the temperature at 50~keV \citep[tens of keV is the typical corona temperature observed in BHXRBs, see e.g.,][]{Yan2020_coronaproperties}, and do \texttt{MONK} simulations to get the theoretical energy and polarimetric spectra. Then we simulate the observed spectra with instrumental responses and fit the spectra in \texttt{XSPEC}. Through spectral parameters, we infer possible coronal geometries in the hard and soft state of BHXRBs. More details about this procedure are presented in the following paragraphs. 

In the simulations, we set the number of photons per geodesic to be 100, making the total number of photons more than $10^7$ in most simulations. Thompson cross section is used. While varying the coronal geometry, the black hole and accretion disk parameters are kept the same. The black hole mass is $10~M_{\odot}$ and the spin is the maximum $a_*=0.998$. The mass accretion rate is 10~\% of the Eddington accretion rate. The inner disk radius is at the innermost stable circular orbit (ISCO) and the outer disk radius is $400~R_{\rm{g}}$ ($R_{\rm{g}}$ is the gravitational radius defined by $GM/c^2$, where $G$ is the gravitational constant, $M$ is the black hole mass, and $c$ is the speed of light), with zero torque at the inner boundary. The spectral hardening factor is fixed at 1.7. The limb darkening and polarization of disk photons follow Chandrasekhar's formula \citep{Chandrasekhar1960} for a semi-infinite scattering atmosphere. The reflection \citep{Dauser2016_reflectionreview} of corona photons and the returning radiation \citep{Dause2022_returningradiation,Mirzaev2024_returingradiation} of disk photons are not included in the simulations.

In practice, to make the simulated energy spectra with \texttt{MONK} comparable with real observations, the \texttt{fakeit}\footnote{\url{https://heasarc.gsfc.nasa.gov/docs/xanadu/xspec/manual/XSfakeit.html}} tool is used to simulate the observed spectra with \textit{NuSTAR} instrumental response and statistical noise for one detector\footnote{\url{https://www.nustar.caltech.edu/page/response-files}: point\_60arcsecRad\_5arcminOA.arf, nustar.rmf, and bgd\_60arcsec.pha
}. The exposure time of the \textit{NuSTAR} simulations is 30~ks. We assume that the source is at 10~kpc and the inclination angle of the disk is $60^\circ$ (in practice, photons between $58^\circ$ and $62^\circ$ are collected). 

We fit the simulated \textit{NuSTAR} spectra in the 3--70~keV energy range with the disk-corona model \texttt{simplcut*kerrbb}, using \texttt{XSPEC} v12.13.0 \citep{Xspec_Arnaud}. \texttt{kerrbb} is the thermal emission model for a thin accretion disk around a Kerr black hole \citep{Li2005_kerrbb}. While \texttt{simplcut} is a uniform scattering model where coronal geometrical effects are not taken into consideration, it is a good enough description for most of the observed cases \citep{Tripathi2021_GX3394,Li2024_EXO1846,Fan2024}, thus enabling us to check which geometries are (not) consistent with what we see in observations. Geometry-dependent corona spectral models are not suitable here because in observations we do not know the coronal geometry beforehand, and we need the power-law index from the fits to distinguish different states empirically. We aim to see from the \texttt{simplcut*kerrbb} fits: i) for which coronal geometries the simulated spectra can(not) be fit well by this simplified spectral model; ii) for the spectra that can be fit well, whether they are more likely in the hard or soft state, judging from the corona parameters and the strength of the disk emission; iii) for the spectra that cannot be fit well, search for any systematic features in the fit residuals and check the geometrical effects.

In the fits, the black hole mass, distance, disk inclination angle and spectral hardening factor in \texttt{kerrbb} are fixed according to the simulation settings, while the spin and mass accretion rate are free. Corona parameters including the photon index $\Gamma$ and the scattering fraction $f_{\rm{sc}}$ are free, while the corona temperature $kT_e$ is fixed at the simulated value 50~keV, due to the poor constraint of this parameter in most of the fits. The ReflFrac parameter in \texttt{simplcut} is set to be 0. The settings of the parameters are summarized in Table~\ref{param_table}. The $\chi^{2}$ statistics is used to find the best-fit values and uncertainties (throughout the paper given at the 90\% of confidence level) of the parameters. 0.5~\% statistical uncertainty is added to the data as a rough estimate of the small fluctuations from Monte Carlo simulations and \textit{NuSTAR} calibration. To derive the disk flux, we calculate the flux of \texttt{$(1-f_{\rm{sc}})$*simplcut*kerrbb} in the 0.1--100~keV range\footnote{\texttt{simplcut*kerrbb} provides the total spectrum, which is the sum of the non-scattered disk photons and the Comptonized photons. The parameter $f_{\rm sc}$ in \texttt{simplcut} describes the fraction of scattered photons, so $(1-f_{\rm{sc}})$ is the fraction of non-scattered disk photons. To calculate the flux of disk photons that are not scattered, we set $f_{\rm sc} = 0$ in \texttt{simplcut} and calculate the flux of $(1-f_{\rm{sc}})$\texttt{*simplcut*kerrbb}.}. We divide the disk flux by the total flux of \texttt{simplcut*kerrbb} in the 0.1--100~keV band to derive the disk flux fraction.

\begin{table}[h!]
\centering
\begin{tabular}{lc}
\hline\hline
& Model\\
& {\tt simplcut*kerrbb}\\
Parameters & \\
\hline\hline
\texttt{simplcut} &\\
$\Gamma$& free\\
$f_{\rm{sc}}$ & free \\
$kT_e$ & 50~keV$^a$ \\
\hline
\texttt{kerrbb}&\\
eta&0\\
$a_*$ & free$^b$ \\
$i$ & $60^{\circ}$~$^c$ \\
$M$ & $10~M_{\odot}$ \\
$D$ & 10~kpc \\
$\dot{M}$ & free \\
hardening factor & 1.7 \\
returning radiation parameter & 0 \\
limb darkening parameter & 1\\
normalization & 1 \\
\hline\hline
\end{tabular}
\caption{Summary of the parameter settings in the fits with the model \texttt{simplcut*kerrbb}. 
$^a$ Except in Section~\ref{temperature_discussion}.
$^b$ Except in Section~\ref{spin}.
$^c$ Except in Section~\ref{inclination}}.
\label{param_table}
\end{table}

Finally, we distinguish different states based on the photon index and the relative strength of the thermal emission. In \cite{Remillard_McClinktock2006ARA}, the fraction of the disk emission is $>75\%$ in the 2--20~keV band in the soft state and $<20\%$ in the hard state. In \cite{Dunn2010}, the criteria is $>80\%$ for the soft state and $<30\%$ for the hard state, based on a global spectral study of 25 BHXRBs. Here, we slightly loosen the standard to be $>70\%$ for the soft state and $<25\%$ for the hard state, taking intermediate states into consideration. Therefore, we distinguish a simulated spectrum with $1.5\leq\Gamma\leq2.0$ and disk fraction $<25\%$ to be in the hard state, and one with $\Gamma>2.0$ and disk fraction $>70\%$ to be in the soft state. Further justification of this criterion will be discussed in Section~\ref{hid}.

Compared with existing works using ray-tracing and Monte Carlo simulation methods to model different coronal geometries \citep[e.g.,][]{Schnittman2010,Zhang2019_MONK,Zhang2022_lamppost_polarization,Urisini_2022}, the key novelty of this work is that we qualitatively compare the simulation results with observations by including instrumental responses and spectral fitting. We explore the parameter space consistent with observations. Meanwhile, compared to a similar ray-tracing code \texttt{kerrC} \citep{Krawczynski2022_kerrC} giving the table model for the wedge and cone corona, the geometries we investigate here are different. Checking these widely adopted coronal geometries and qualitatively distinguishing them in different states is important for both observational and theoretical studies. From the observational perspective, a better knowledge of the proper coronal model for data fitting makes parameter estimation more reliable. From the theoretical perspective, assessing the validity of simplifications in modeling (e.g., the point-like corona assumed in reflection \citep{dauser2014relxilllp} and reverberation \citep{Uttley2014_reverberation} modeling) against observations provides the foundation for applying these models in observations.

We note the readers that the above simulation settings are simplified views drawing from our current understandings of observations. General relativistic magnetohydrodynamic (GRMHD) simulations can reproduce some coronal models \citep{Yuan2014}. For example, some works take the inner hot accretion flow as the corona and the outer cold accretion flow as the disk \citep[e.g., ][]{Naethe2025,Hankla2025}. Other works take the sheath between the jet and wind (the jet base) as the corona \citep[e.g., ][]{Shashank2025,Sridhar2025}. In GRMHD simulations, the corona has a complex temperature and density profile. These are not taken into consideration in our work because the simplified constant density and temperature coronae are already able to reproduce spectra consistent with current observations. Incorporating the GRMHD simulation results into spectral modeling will be an interesting direction to explore in the future, especially when higher-resolution data from future missions become available, to better constrain the physical structure and properties of the corona.

\section{Simulations and Results}\label{result}

\subsection{Disk Emission}\label{kerrbb}

We first simulate a pure disk spectrum without corona scattering to ensure that the result given by \texttt{MONK} is consistent with the \texttt{kerrbb} model. As is mentioned above, in this and subsequent sections the mass accretion rate is set at $10\%~\dot{M}_{\rm{Edd}}$ ($2.23\times10^{18}~\rm{g/s}$ for a $10~M_{\odot}$ black hole) and the spin is 0.998 in the simulations. The hardening factor is set at 1.7, and the limb darkening and polarization of disk photons follow Chandrasekhar's formula \citep{Chandrasekhar1960}. The disk has zero torque at the inner boundary, corresponding to eta = 0. The disk reflection or self-returning radiation is not considered. For the parameters of \texttt{kerrbb} in the fits, see Table~\ref{param_table}. The fitting result is shown in Fig.~\ref{kerrbb}, with $\dot{M}=(2.23\pm0.04)\times10^{18}~\rm{g/s}$, $a_*=0.997\pm0.001$ and $\chi^2/d.o.f.=135.23/124=1.09$ ($d.o.f.$ refers to degree of freedom), showing that the disk spectrum from the \texttt{MONK} simulation is consistent with that given by \texttt{kerrbb}. Therefore, we can confidently use \texttt{kerrbb} as the seed disk spectrum and study the impact of Compton scattering on the seed photons.

\begin{figure}
    \centering
    \includegraphics[width=0.95\linewidth]{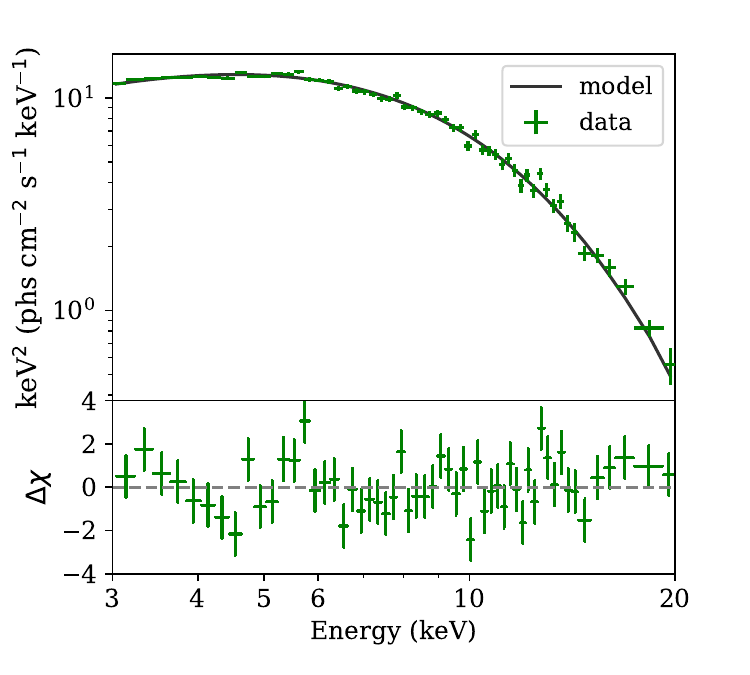}
    \caption{The data, best-fit model and residuals of the fit of the simulated disk emission of a $10~M_{\odot}$, $a_*=0.998$ black hole accreting at 10\% of the Eddington mass accretion rate, observed from 10~kpc and $60^{\circ}$. The model \texttt{kerrbb} can fit well the simulated spectrum, with $\chi^2/d.o.f.=135.23/124=1.09$.}
    \label{kerrbb}
\end{figure}

\subsection{Sandwich Geometry}\label{sandwich}

\begin{figure*}[htbp]
    \centering
    \includegraphics[width=0.44\linewidth]{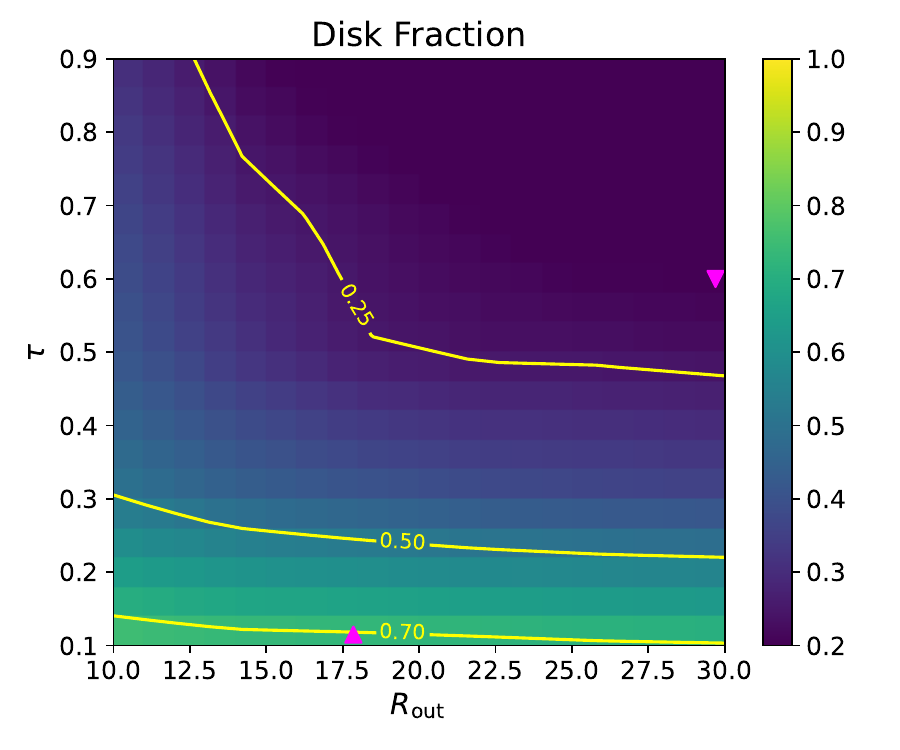}
    \includegraphics[width=0.44\linewidth]{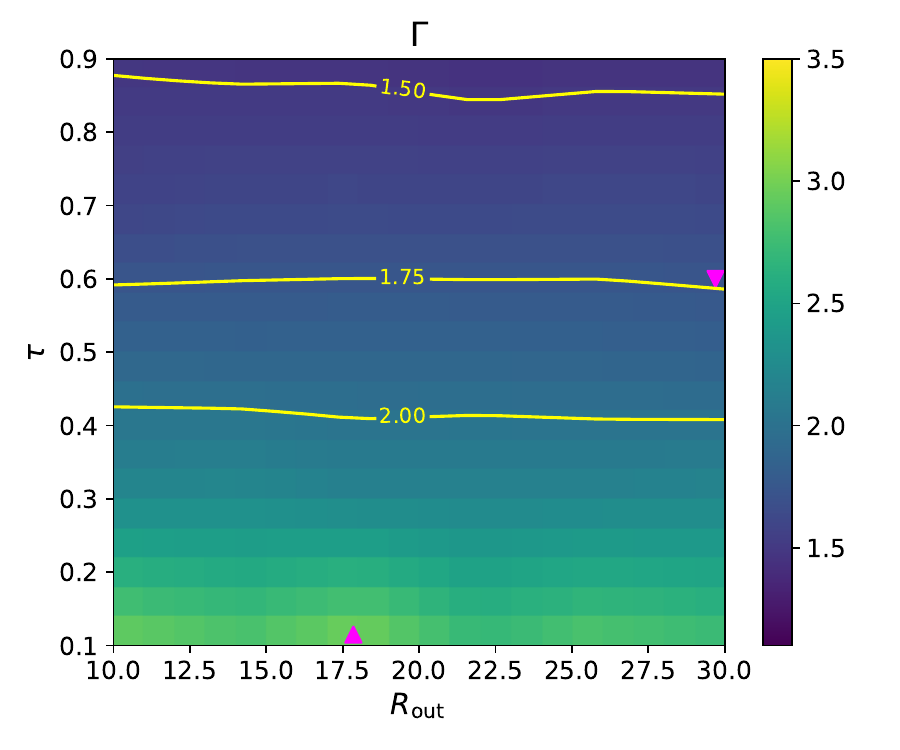}
    \caption{Sandwich corona. The change of the disk fraction (the left panel) and $\Gamma$ (the right panel) with the optical depth ($\tau=n_{\rm{e}}\sigma_{\rm{T}}h$, where  $h=h_{\rm{max}}-h_{\rm{min}}$ is the thickness of the corona) and outer corona radius ($R_{\rm{out}}$), observed at $i=60^{\circ}$. We produce the simulation over a $6 \times 5$ grid ($\tau=0.1,0.3,0.5,0.7,0.9$ while $R_{\rm{out}}=10, 14, 18, 22, 26, 30~R_{\rm{g}}$) and perform a 2D interpolation to get the continuous trend and draw the contour lines. The upper triangle and lower triangle show the selected soft-state and hard-state spectra of the sandwich corona in Fig.~\ref{pol_compare}. Throughout the paper, the color bars of all the contour plots are the same.}
    \label{sandwich_grid}
\end{figure*}

In the sandwich geometry, the corona is two parallel layers with a certain thickness above and below the disk. We set the minimum height of the corona above (below) the disk at $1~R_{\rm{g}}$, the maximum height at $2~R_{\rm{g}}$, and the inner corona radius at the ISCO. We simulate the sandwich corona with different outer radii $R_{\rm out}$ and vertical optical depths $\tau$ (defined as $\tau=n_{\rm{e}}\sigma_{\rm{T}}h$, where $h=h_{\rm{max}}-h_{\rm{min}}$ is the thickness of the corona). Note that for high inclinations, the effective optical depth can be significantly larger than the vertical $\tau$.  We do the simulations on a $6 \times 5$ grid ($\tau=0.1,0.3,0.5,0.7,0.9$ while $R_{\rm{out}}=10, 14, 18, 22, 26, 30~R_{\rm{g}}$) and do a 2D interpolation to get the continuous trend of the disk fraction and $\Gamma$. The results are shown in Fig.~\ref{sandwich_grid}.

All the spectra in the parameter space of our simulations can be fit with $\chi^2/d.o.f. < 2$. $\Gamma$ can be well constrained with an uncertainty of $\sigma_{\Gamma} < 0.2$ at the 90\% confidence level. (Because the constraints on $\Gamma$ are weak at smaller optical depths, we omit this part from the plot.) From Fig.~\ref{sandwich_grid}, more photons are scattered when the corona outer radius is larger and the optical depth is larger, therefore the disk fraction, i.e., the X-ray flux fraction of the observed disk brightness compared to the total flux, is lower and the Comptonized photons are more dominant. The photon index $\Gamma$ is mainly determined by the corona optical depth, not the outer radius of the corona. This indicates that the non-uniform covering of the corona over the disk does not significantly influence the energy redistribution of scattered photons, possibly because the majority of disk photons come from the inner disk. $\Gamma$ becomes lower with the increase of the optical depth, because the increasing number of lower energy photons scattered to higher energy contributes to a higher power-law tail.

As indicated by the contour lines on Fig.~\ref{sandwich_grid}, for the sandwich geometry, the lower part of the parameter space (optical depth smaller than $\sim 0.15$) is the possible model for the soft state, while the upper-right part of the parameter space (corona radius larger than $\sim 18~R_{\rm{g}}$, and optical depth larger than $\sim 0.5$ and smaller than $\sim 0.9$) is the possible model for the hard state. Note that due to the uncertainties of the fitting parameters and the calculation of the disk fraction, the boundary of the soft and hard state in the parameter space is also an estimation instead of a strict delineation.

To test the impact of the thickness and height of the sandwich corona, we choose the largest radius and optical depth in our simulation grid ($\tau=0.9$ and $R_{\rm{out}}=30~R_{\rm{g}}$), because the higher probability of scattering of this case should illustrate better the influence of the coronal height and thickness. Fixing the thickness at $1~R_{\rm{g}}$, we test a corona close to the disk ($h_{\rm{min}}=0$) and a corona well above the disk compared with the size of the black hole ($h_{\rm{min}}=2$ and $5~R_{\rm{g}}$). Fixing the height at $1~R_{\rm{g}}$, we test coronae with different thicknesses ($h_{\rm{max}}-h_{\rm{min}}=0.5,~2$ and $5~R_{\rm{g}}$). Compared with the result shown above, these changes of the thickness and height of the corona change $\Gamma$ only $\lesssim0.03$ (which is just around the fitting uncertainty of $\Gamma$) and the disk fraction $\lesssim5\%$. Our results thus indicate that the height and thickness of the corona play a minor role in deciding the spectral shape, which is expected because the key in the physical picture of the sandwich corona is the constant optical depth covering a certain fraction of the disk. Neglecting light-bending effects, the effective optical depth does not change with the coronal height or thickness. When light-bending effects are taken into consideration, the track of the photon traveling in the corona will be influenced, but the impact turns out to be considerably small. Further larger $h_{\rm{min}}$ is not within our interest because we are considering a scattering plasma covering the disk instead of a hot electron cloud far above the disk.

\subsection{Spherical Geometry}\label{sphere}

\begin{figure*}[htbp]
    \centering
    \includegraphics[width=0.44\linewidth]{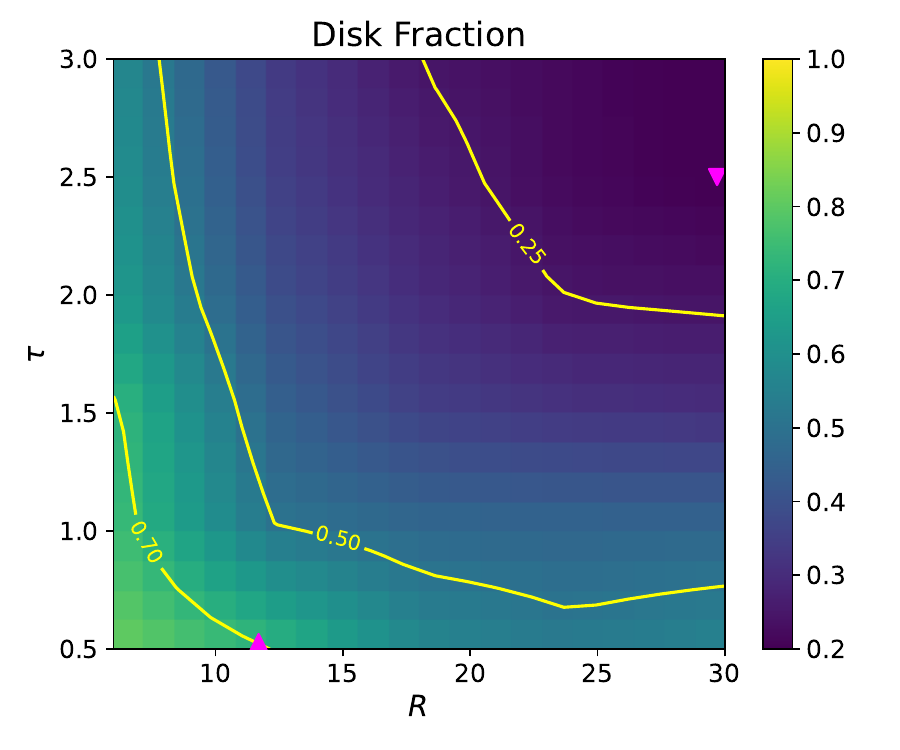}
    \includegraphics[width=0.44\linewidth]{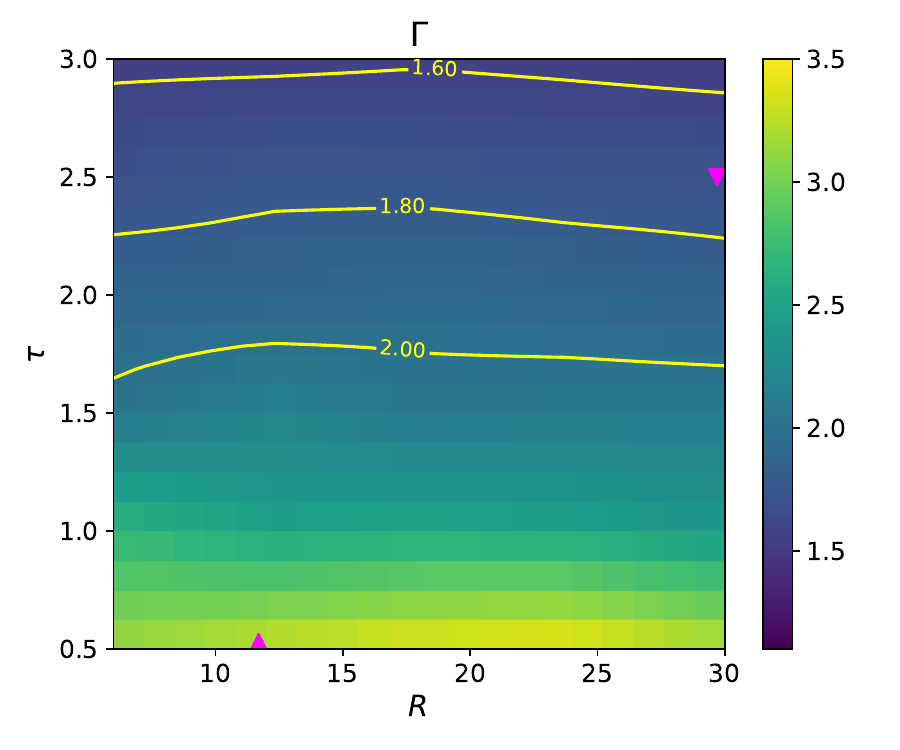}
    \caption{Spherical corona. The change of the disk fraction (the left panel) and $\Gamma$ (the right panel) with the optical depth ($\tau=n_{\rm{e}}\sigma_{\rm{T}}R$) and corona radius ($R$), observed at $i=60^{\circ}$. We produce the simulations over a $5 \times 5$ grid ($\tau=1.0,1.5,2.0,2.5,3.0$ while $R=6, 12, 18, 24, 30~R_{\rm{g}}$) and perform a 2D interpolation to get the continuous trend and draw the contour lines. The upper triangle and lower triangle show the selected soft-state and hard-state spectra of the spherical corona in Fig.~\ref{pol_compare}.}
    \label{sphere_grid}
\end{figure*}

In the spherical geometry, the corona is a central sphere around the black hole, covering part of the inner disk. We simulate spherical coronae with varying radii $R$ and optical depths $\tau$ (defined as $\tau=n_{\rm{e}}\sigma_{\rm{T}}R$, where the corona radius $R$ is defined from the center of the corona). We do the simulations on a $6 \times 5$ grid ($\tau=0.5,1.0,1.5,2.0,2.5,3.0$ while $R=6, 12, 18, 24, 30~R_{\rm{g}}$). The results are shown in Fig.~\ref{sphere_grid}.

All the spectra in the parameter space of our simulations can be fit with $\chi^2/d.o.f. < 2$. $\Gamma$ can be constrained well with $\sigma_{\Gamma}\leq 0.2$. Similar to the sandwich corona, $\Gamma$ decreases with the increase of optical depth, and does not strongly depend on the corona radius.
The disk fraction decreases with the increase of optical depth and corona radius.

As indicated by the contour lines on Fig.~\ref{sphere_grid}, for the spherical geometry, the lower left part of the parameter space (below the contour line of 70\% disk fraction, i.e., the triangle-like region bounded by $\tau=1.5, R=6~R_{\rm{g}}$ and $\tau=0.5, R=12~R_{\rm{g}}$, and the line connecting those two points) is the possible model for the soft state, while the upper right part of the parameter space (optical depth larger than $\sim 2$ and smaller than $\sim 3$, and radius larger than $\sim 18~R_{\rm{g}}$) is the possible model for the hard state.

\subsection{Lamppost Geometry}\label{lamppost}

\begin{figure*}
    \centering
    \includegraphics[width=0.44\linewidth]{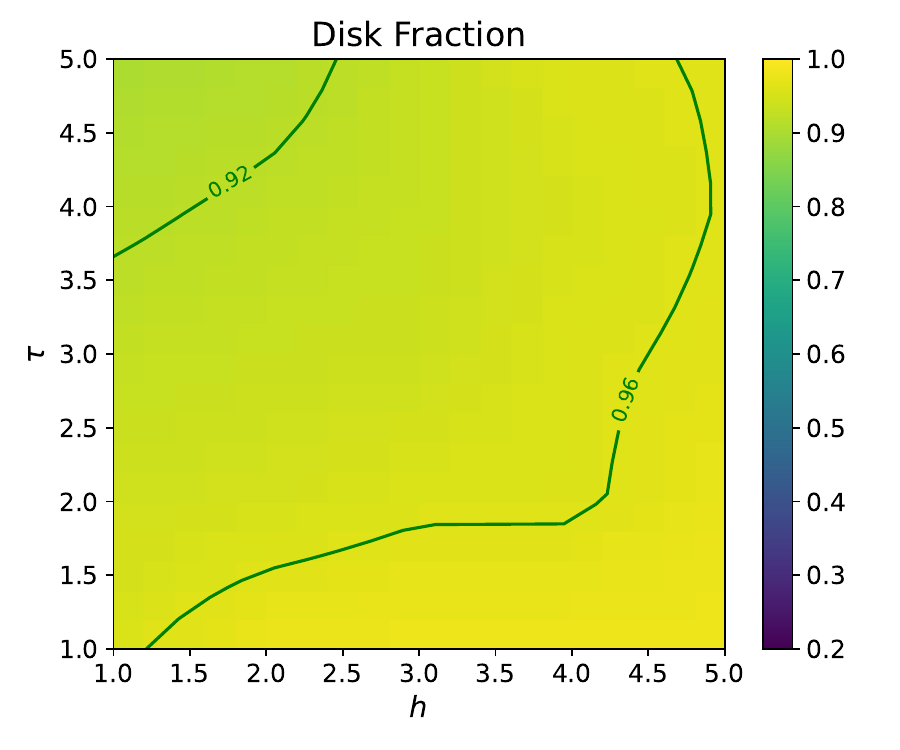}
    \includegraphics[width=0.44\linewidth]{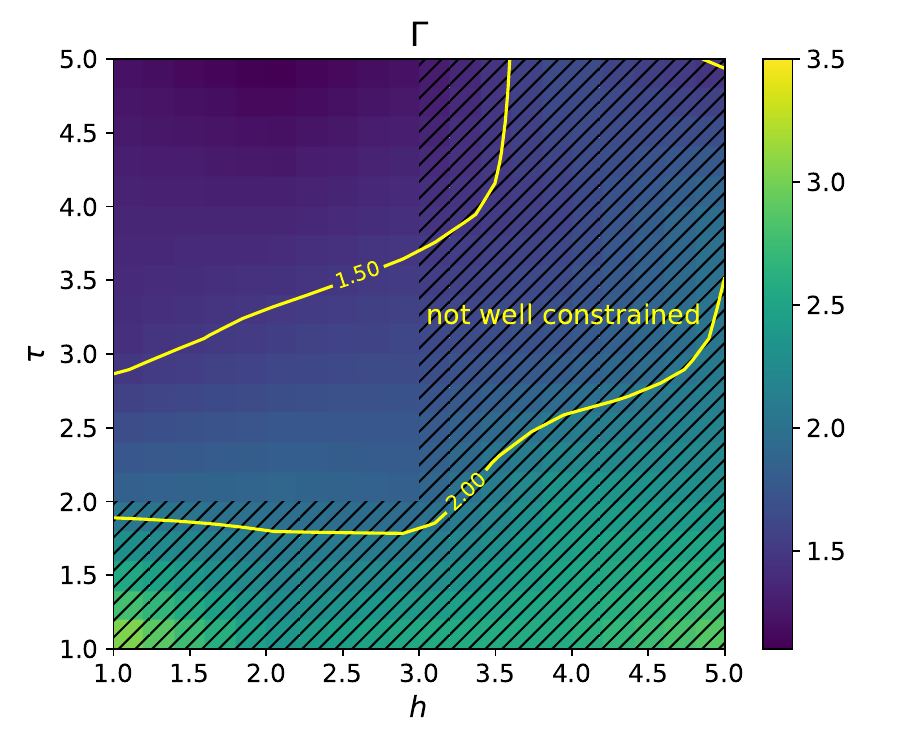}
\caption{Lamppost corona. The change of the disk fraction (the left panel) and $\Gamma$ (the right panel) with the optical depth ($\tau=n_{\rm{e}}\sigma_{\rm{T}}R$, where $R=2~R_{\rm{g}}$) and corona height ($h$) above the disk, observed at $i=60^{\circ}$. We do the simulation on a $5 \times 5$ grid ($\tau=1,2,3,4,5$ while $h=1,2,3,4,5~R_{\rm{g}}$) and do a 2D interpolation to get the continuous trend and draw the contour lines. The shadowed region indicates the spectra with $\sigma_{\Gamma}$ larger than 0.2.}
    \label{lamppost_grid}
\end{figure*}

In the lamppost geometry, the corona is a point-like source along the black hole spin axis. In our simulations, to ensure that some photons are scattered, we use a finite size spherical corona ($R=2~R_{\rm{g}}$) to represent the lamppost corona. We simulate the lamppost corona with varying heights $h$ and optical depths $\tau$ (defined as $\tau=n_{\rm{e}}\sigma_{\rm{T}}R$, where $R=2R_{\rm{g}}$ is the radius of the corona). We do the simulations on a $5 \times 5$ grid ($\tau=1,2,3,4,5$ while $h=1,2,3,4,5~R_{\rm{g}}$). The results are shown in Fig.~\ref{lamppost_grid}.

From Fig.~\ref{lamppost_grid}, the disk emission is always dominant even if the corona is already very close to the black hole and disk. Due to the limited size of the corona and light bending effects, the majority of the photons do not reach the corona and are not scattered. Because of the dominance of non-scattered photons, the photon index cannot be constrained well (with uncertainty $\sigma_{\Gamma}$ larger than 0.2, sometimes even reaching the parameter limit) when the corona is high (higher than $\sim3R_{\rm{g}}$) above the disk or the optical depth is low (lower than $\sim2$). These can possibly be the case of the soft state. However, when $\Gamma$ is constrained below 2, the disk fraction is still larger than 90\%, which is not consistent with the hard state spectrum. If we increase the coronal height, the disk fraction increases and we cannot constrain $\Gamma$, which means that such a configuration cannot describe the hard state and is therefore not meaningful to explore it further.

We also test the impact of the value of the coronal radius. For a coronal radius in the range $R=1~R_{\rm{g}}$ to $R=5~R_{\rm{g}}$, there are no significant variations in the value of $\Gamma$ with respect to the case $R=2~R{\rm{g}}$. The disk fraction for $R=1~R_{\rm{g}}$ is always higher than 90\% and that for $R=5~R_{\rm{g}}$ is always higher than 55\%, which is still too disk-dominated to describe the hard state. By definition (and in the simplifications used in reverberation modeling) the lamppost corona is a point-like source along the black hole spin axis. Our aim is to test how well the point-like simplification can work. Even if the case $R=5~R_{\rm{g}}$ is already an extended corona, we cannot reproduce a hard state. While the lamppost coronal geometry is a useful simplification in reverberation analysis and reflection modeling, it may not be a very realistic scenario when the spectrum is hard and dominated by the coronal emission. It is also difficult to explain the polarization observation results in \cite{Gianolli2023,Krawczynski2022} with this geometry. While a further extended a harder spectrum may be possible for extended or hybrid lamppost-like geometries, these cases are beyond our interest.

\section{discussion}\label{discussion}
\subsection{Consistency with \texttt{compTT}}\label{CompTT}

While most of the corona models commonly used in observational studies do not assume a certain coronal geometry, there are a few ones taking geometrical effects into consideration analytically or numerically, for example, \texttt{compTT} \citep{Titarchuk1994_CompTT,Hua_Titarchuk1995_CompTT} and \texttt{compPS} \citep{Poutanen_Svensson_1996_CompST}. Here, as a benchmark, we compare our Monte Carlo simulation and fitting results with the \texttt{compTT} model to check their consistency.

The \texttt{compTT} model is an additive model describing Comptonization of soft photons in a hot plasma, using analytical approximation to model the dependence of the optical depth on coronal geometry \citep[disk-like or spherical corona,][]{Titarchuk1994_CompTT}. The input thermal spectrum is a Wien law. The model parameters include the redshift, the input soft photon temperature, the plasma temperature and optical depth, a geometrical parameter, and the normalization factor.

Similar to how we fit the simulated \texttt{MONK} spectra, we generate the \texttt{compTT} model spectra of different corona optical depths and geometries, simulate the observed spectra with \textit{NuSTAR} instrumental effects, and fit them with \texttt{simplcut*diskbb} \citep[\texttt{diskbb} is a non-relativistic multi-temperature blackbody disk model,][]{Mitsuda}. The plasma temperature in the model and fits is fixed at 50~keV. The temperature of thermal photons is set at 1~keV in \texttt{compTT}, and it is a free fitting parameter in \texttt{diskbb}.

\begin{figure*}
    \centering
    \includegraphics[width=0.43\linewidth]{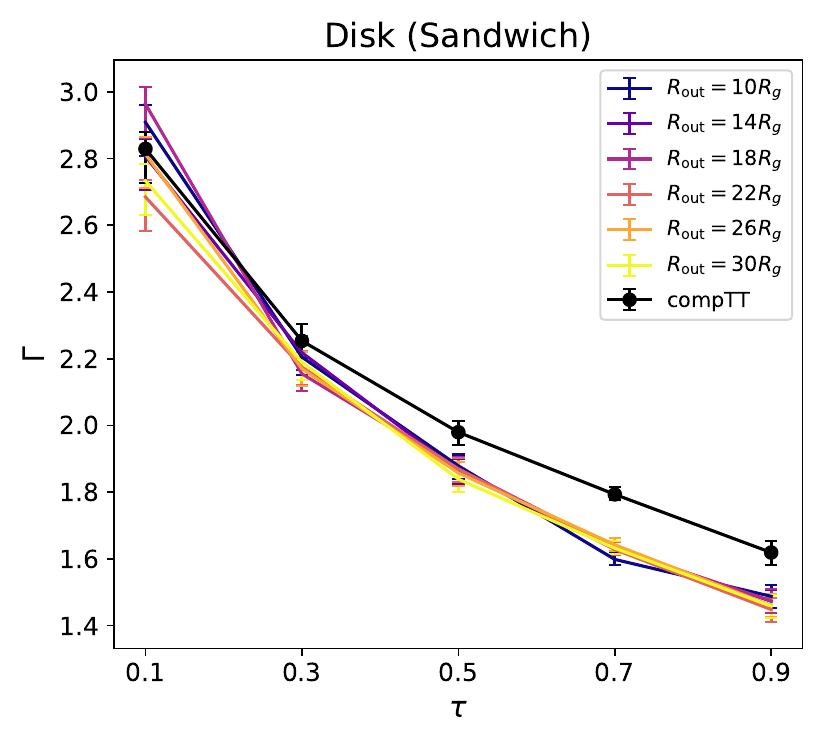}
    \includegraphics[width=0.43\linewidth]{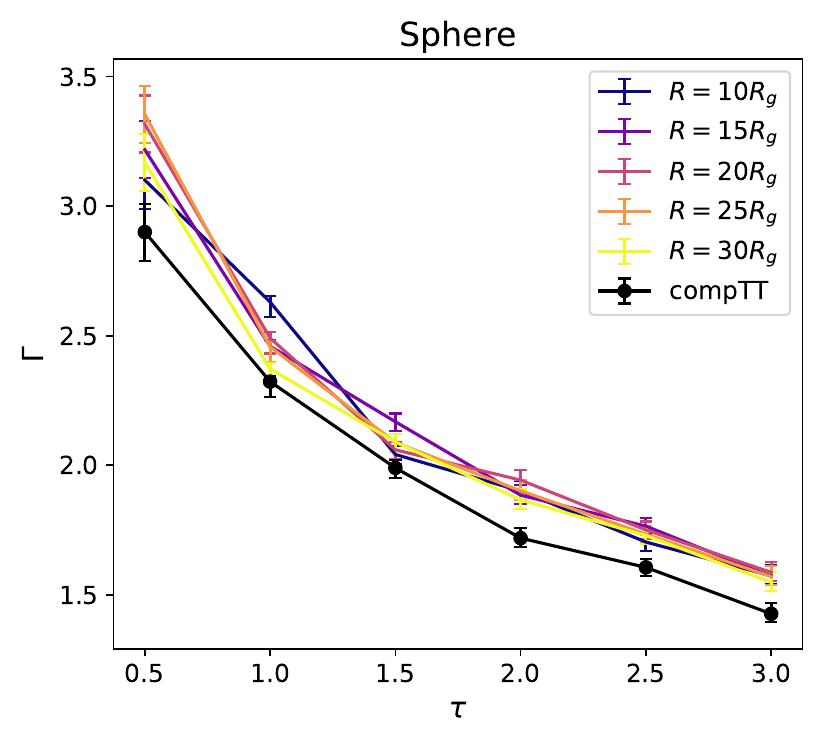}
    \caption{The change of $\Gamma$ with corona optical depth given by fitting the \texttt{compTT} model and \texttt{MONK} simulation spectra with \texttt{simplcut*diskbb}. The left panel shows the result for a disk-like (sandwich) corona and the right panel shows the result for a spherical corona. Results of different sandwich corona outer radii and spherical corona radii are shown in different colors. While the plots show the general consistency of the changing trend of $\Gamma$ given by \texttt{MONK} and \texttt{compTT}, the absolute values of $\Gamma$ show discrepancies depending on the size of the corona.}
    \label{compTT}
\end{figure*}

Fig.~\ref{compTT} shows how $\Gamma$ given by the \texttt{compTT} model and our simulations changes with the optical depth in the disk-like and spherical corona, respectively. The changing trend of $\Gamma$ with the optical depth given by the two models is similar. However, when the optical depth is higher and $\Gamma$ can be constrained better, the Monte Carlo simulation results start to deviate from \texttt{compTT}. For $\tau\gtrsim0.5$ in the sandwich geometry, the \texttt{compTT} value is higher than the \texttt{MONK} ones. For $\tau\gtrsim2.0$ in the spherical geometry, the \texttt{compTT} value is lower than the \texttt{MONK} ones. These deviations can be caused by several simplifications in \texttt{compTT}: i) though the disk and spherical geometries are distinguished in the \texttt{compTT} model, it is an analytical approximation, only taking different boundary conditions into account when the scattering number distribution of photons is calculated; ii) light bending and gravitational redshift effects are not considered; iii) the input thermal spectrum is a Wien law instead of a realistic disk model with a radial temperature profile. The comparison shows the necessity to take these effects into account in strong scattering regime to get more accurate modeling of the observed spectra.

\subsection{Different states on the HID}\label{hid}

\begin{figure*}
    \centering
    \includegraphics[width=0.47\linewidth]{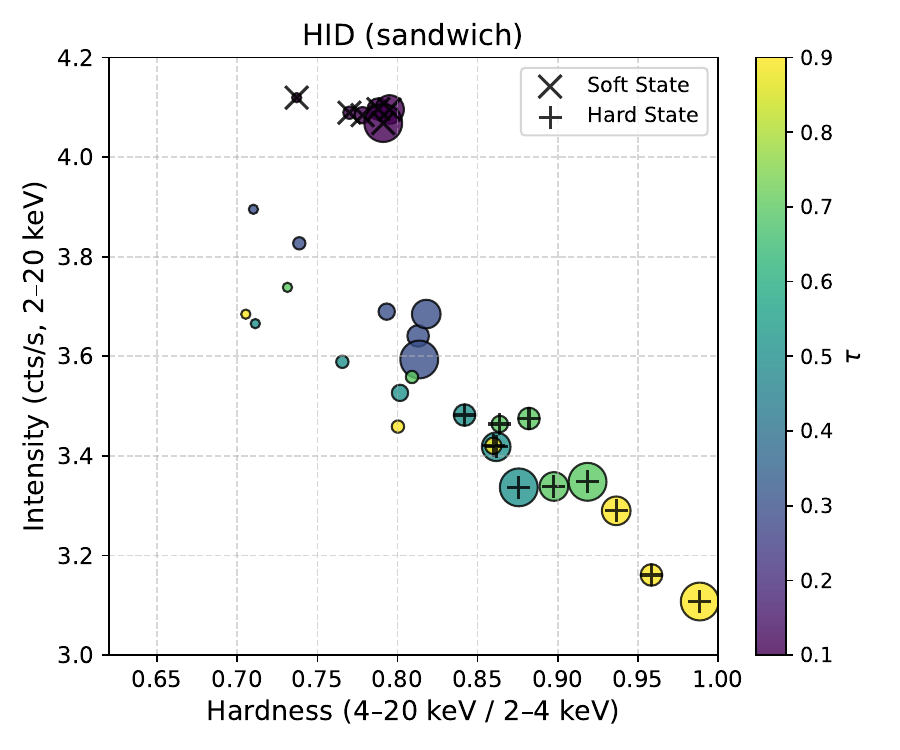}
    \includegraphics[width=0.47\linewidth]{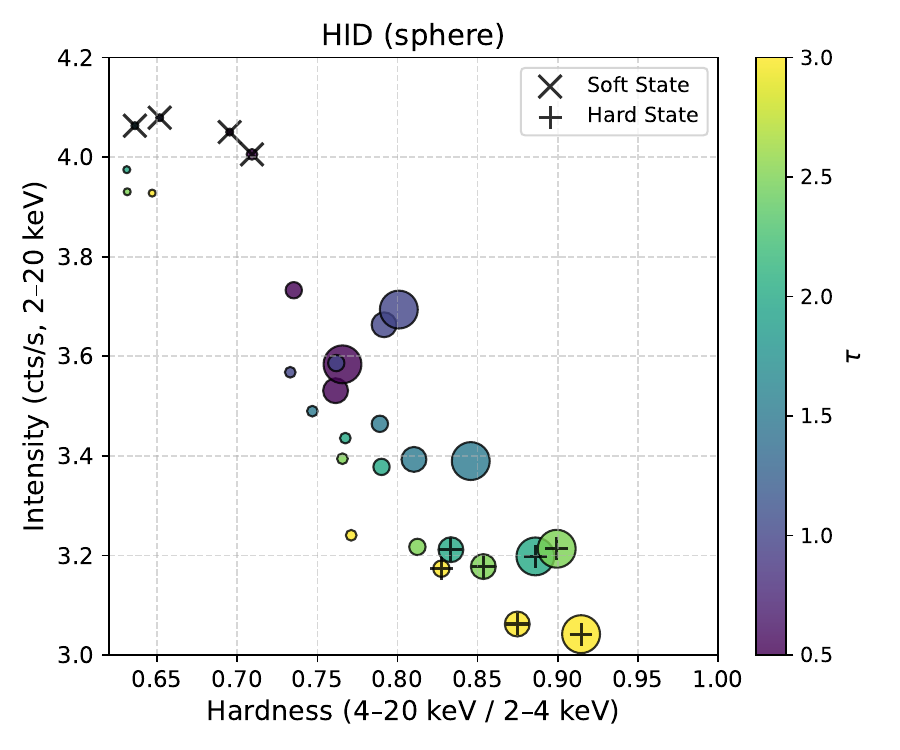}
    \caption{The HID of the simulated points of the sandwich corona (left panel) and spherical corona (right panel). The $x$-axis is the hardness ratio of the count rate of photons between 4--20~keV to 2--4~keV, and the $y$-axis is that of photons between 2--20~keV. The size of the point shows the size of the coronae (rescaled), and the color reflects the optical depth. The crosses note the points identified as the soft state in Section~\ref{result}, while the plus signs note the hard state points.}
    \label{hid_plot}
\end{figure*}

The hardness-intensity diagram (HID) is a commonly used way to distinguish different states in observations. However, because the absolute value of hardness and count rate depend on the instrument used, we adopt the disk fraction as the criterion in the above analysis. Here, we double check this method by plotting the HID of the simulated points in Fig.~\ref{hid_plot}. We can see that, for both the sandwich and spherical corona, the points identified as the soft state appear on the upper left corner of the HID, while the hard-state points appear on the lower right corner. This indicates the consistency of two different criteria in distinguishing different states.

\subsection{Impact of the inclination angle}\label{inclination}
\begin{figure*}
    \centering
    \includegraphics[width=0.85\linewidth]{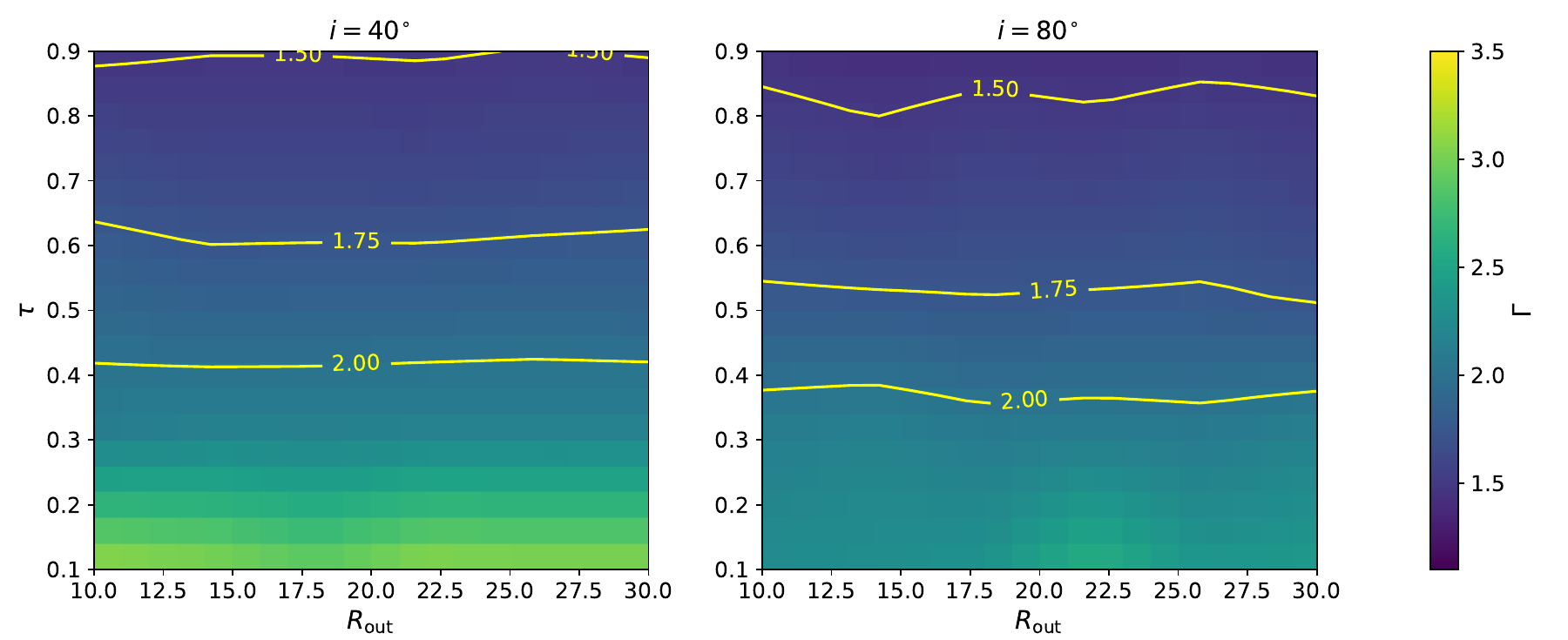}
    \includegraphics[width=0.85\linewidth]{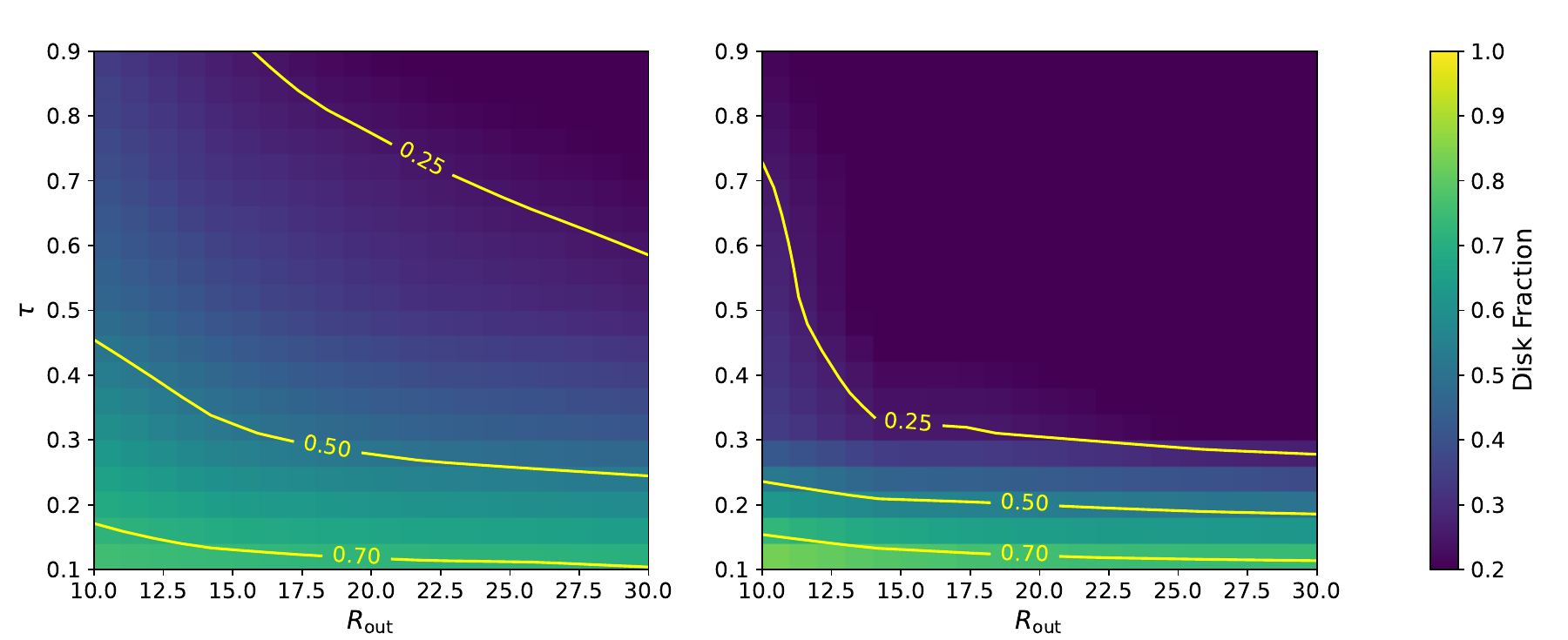}
    \caption{Sandwich corona with different disk inclinations. The change of the photon index (the upper panels) and disk fraction (the lower panels) with the optical depth and outer radius, observed at $i=40^{\circ}$ (the left panels) and $i=80^{\circ}$ (the right panels). The simulation grid is the same as that in Fig.~\ref{sandwich_grid}.}
    \label{incl_compare}
\end{figure*}

In the previous section, we assumed that the disk inclination angle is $60^{\circ}$. To further  investigate the impact of the disk inclination angle on our results, we calculate the spectra of the sandwich corona at a low inclination angle ($40^{\circ}$) and a high inclination angle ($80^{\circ}$). The spectral fitting results are shown in Fig.~\ref{incl_compare}.

The changing trend of $\Gamma$ and disk fraction with the optical depth and outer corona radius is still similar to the results for a medium inclination angle in Fig.~\ref{sandwich_grid}. $\Gamma$ decreases with the increase of $\tau$, and the disk fraction decreases with the increase of $\tau$ and $R_{\rm{out}}$. When the inclination angle is higher, $\Gamma$ and disk fraction is slightly lower for the same $\tau$ and $R_{\rm{out}}$. While the difference in the inclination angle does not change our general understanding of the corona properties of different states, it affects the boundary line between states.

A qualitative explanation for the differences in $\Gamma$ and disk fraction can be that the effective optical depth of a photon collected at $i$ should be $h/\cos i$ (where $h$ is the thickness of the sandwich corona and $i$ is the inclination angle), which increases with the increase of $i$. However, this is just a rough estimation without considering the relativistic light bending effects and the averaging of photons from different parts of the disk.

\subsection{Impact of the spin value}\label{spin}

\begin{figure*}
    \centering
    \includegraphics[width=0.44\linewidth]{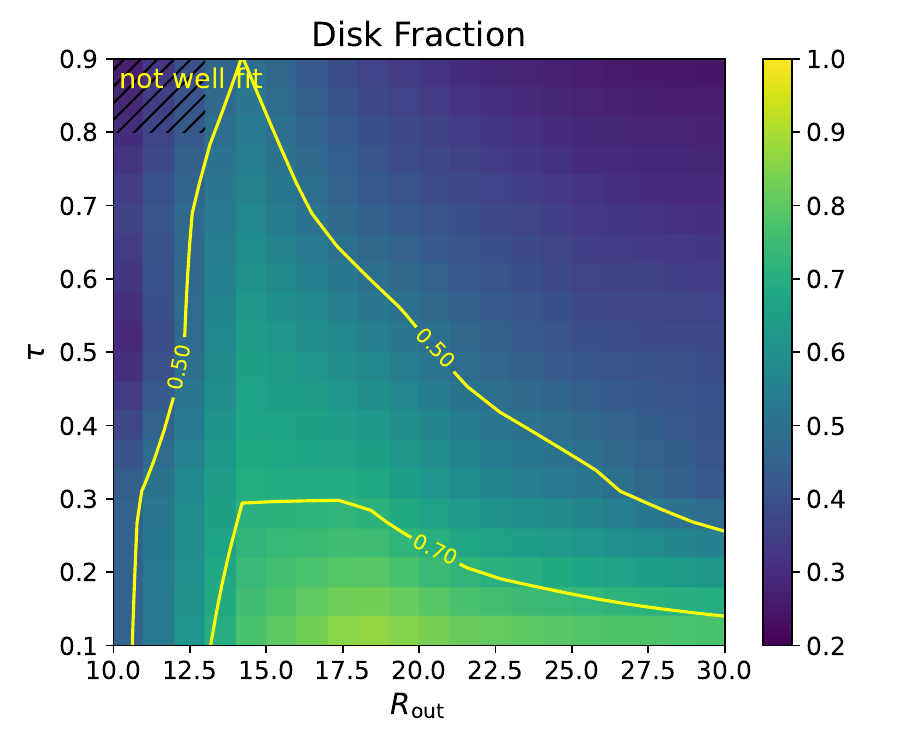}
    \includegraphics[width=0.44\linewidth]{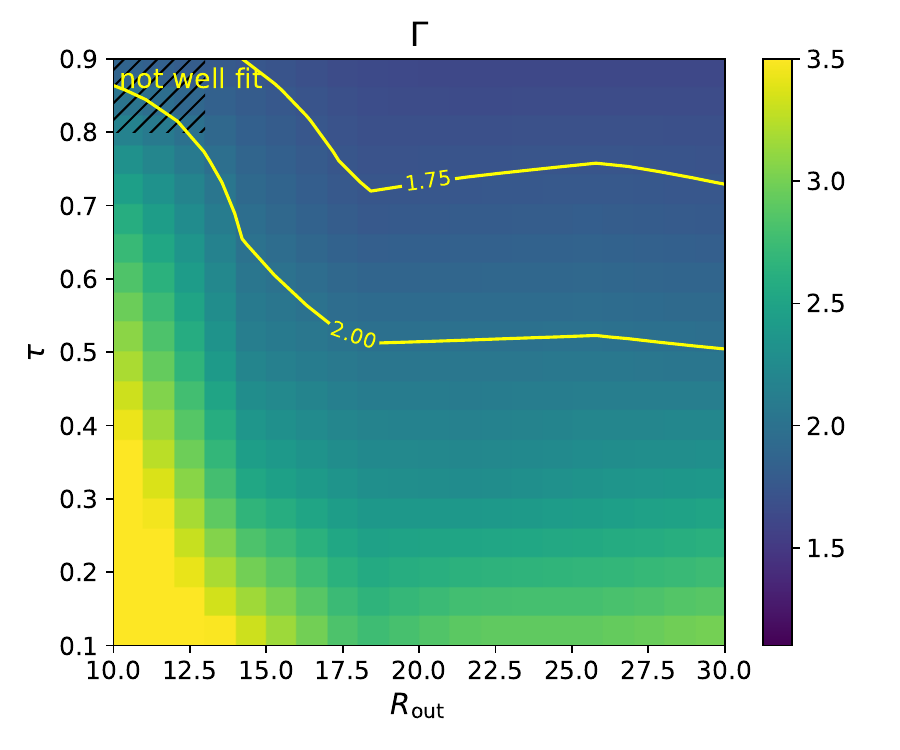}

    \includegraphics[width=0.44\linewidth]{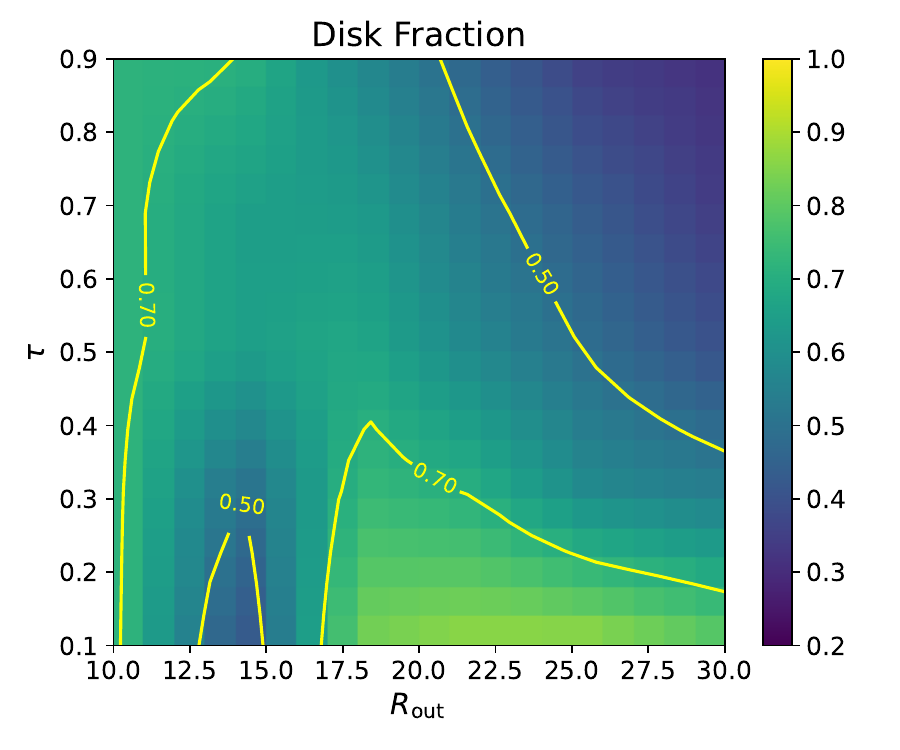}
    \includegraphics[width=0.44\linewidth]{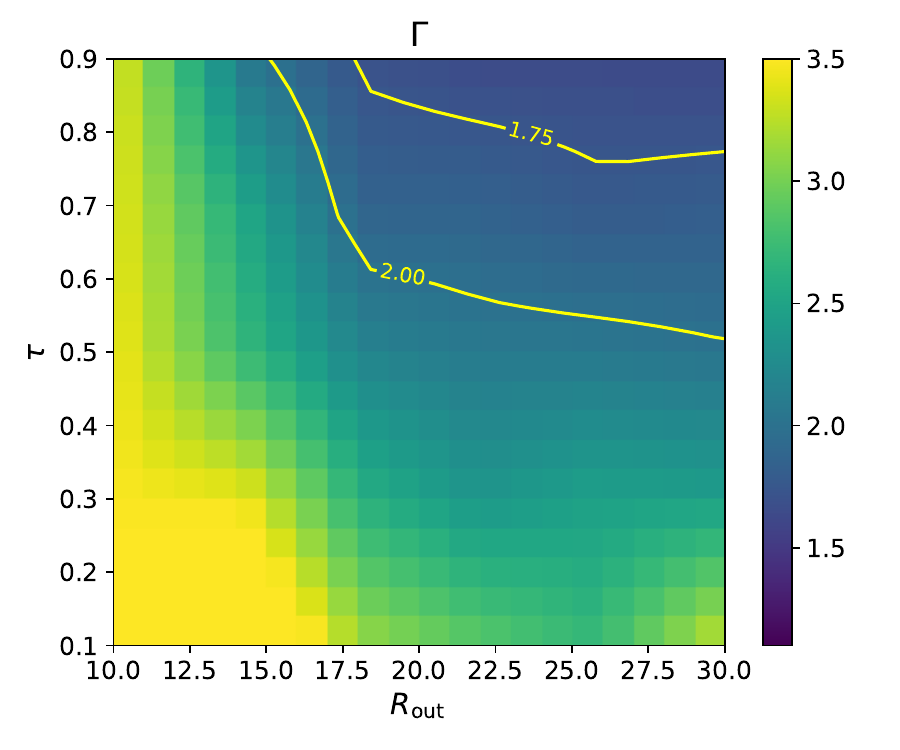}
    
    \caption{Sandwich corona with the black hole spin $a_*=0.5$ (the upper panels) and $a_*=0$ (the lower panels). The change of the disk fraction (the left panels) and the photon index (the right panels) with the optical depth and outer radius, observed at $i=60^{\circ}$. The simulation grid is the same as that in Fig.~\ref{sandwich_grid}. The shadowed region indicate the spectra cannot be fit well with $\chi^2/d.o.f.>2$.}
    \label{sandwich05_grid}
\end{figure*}

\begin{figure}
    \centering
    \includegraphics[width=0.9\linewidth]{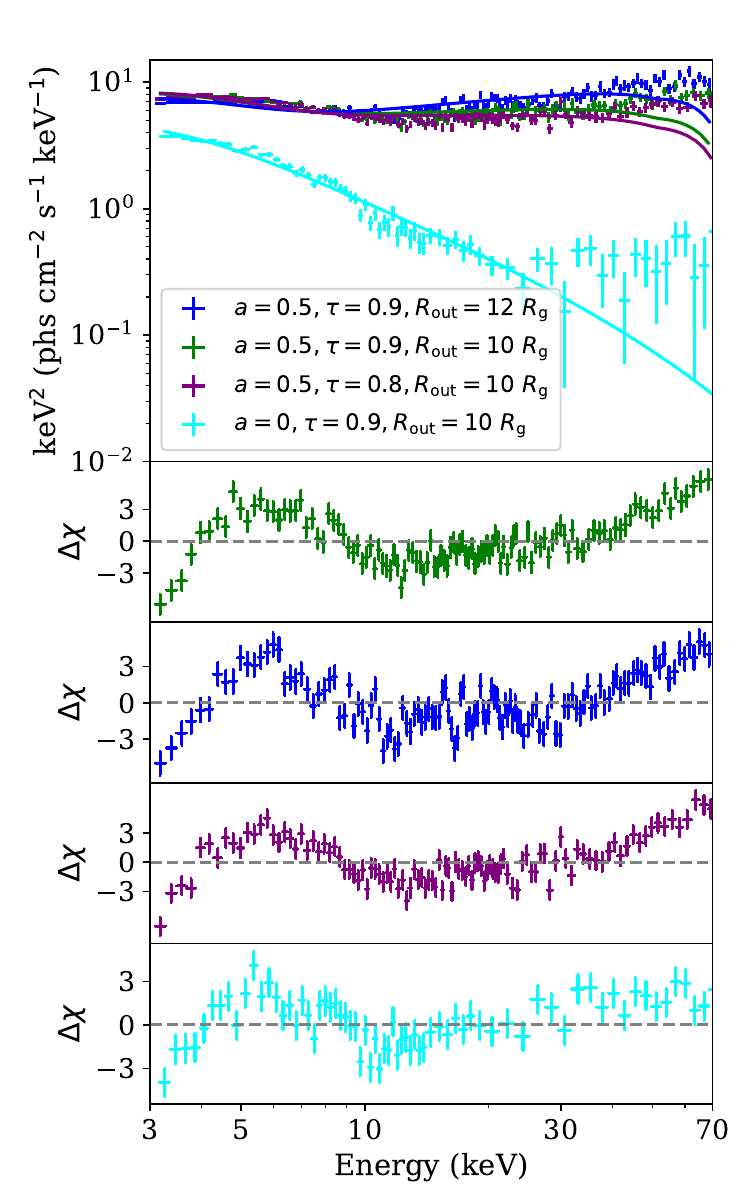}
    \caption{Four example spectra that cannot be fit well by the \texttt{simplcut*kerrbb} model. They are the spectra of a $a_*=0.5$ or $a_*=0$ black hole with a sandwich corona, with inner radius equals the ISCO and different outer radii and optical depths noted in the labels. The free parameters in these fits are the mass accretion rate, scattering fraction and $\Gamma$.}
    \label{sandwich05_spec}
\end{figure}

To study the impact of the spin of the black hole on the corona spectral properties, we repeat the simulations of the sandwich corona around a $a_*=0.5$ and a $a_*=0$ black hole. In Boyer-Lindquist coordinates, the ISCO of the $a_*=0.5$ black hole is $4.23~R_{\rm{g}}$ and that of the $a_*=0$ black hole is $6~R_{\rm{g}}$. The inner radius of the sandwich corona is also set to be the ISCO of the disk. This means that compared with the setting in subsection \ref{sandwich}, the corona with the same outer radius covers less fraction of the disk around a black hole with a lower value of the spin parameter.

In the fits, we find that the mass accretion rate and spin are highly degenerate, especially when the Comptonization component is stronger. Therefore, to reduce the influence of the fluctuation of the disk parameters and focus on the impact of the spin on the corona parameters, we freeze the spin to the set value 0.5 or 0 and leave the mass accretion rate free in the fit. The fitting results are shown in Fig.~\ref{sandwich05_grid}. For the $a_*=0.5$ black hole, when the outer radius of the corona is larger than $\sim 15~R_{\rm{g}}$, the trend of the disk fraction and $\Gamma$ is similar to that in Fig.~\ref{sandwich_grid}. For the $a_*=0$ black hole, the critical radius is $\sim 18~R_{\rm{g}}$. For smaller outer corona radii, the changing trend of disk fraction and $\Gamma$ is different from what we see for a $a_*=0.998$ black hole, i.e, the disk fraction or $\Gamma$ do not decrease with the increasing coronal radius and optical depth. When the outer corona radius is smaller than $\sim12~R_{\rm{g}}$ and $\tau$ is larger than $\sim0.8$ around a $a_*=0.5$ black hole, the \texttt{simplcut*kerrbb} model cannot fit well the simulated spectra with $\chi^2/d.o.f. > 2$. A few examples of these spectra are shown in Fig.~\ref{sandwich05_spec}, showing that the thermal emission is distorted and the power-law tail is underestimated in the fits. Although statistically the spectra in the similar parameter space for the $a_*=0$ black hole can be fit with $\chi^2/d.o.f.$ slightly lower than 2, the main reason for that may be the reduced luminosity of a lower spin black hole and larger uncertainties in observational data. As is shown in Fig.~\ref{sandwich05_spec}, the residuals for a $\tau=0.9$, $R_{\rm{out}}=10~R_{\rm{g}}$ corona around the $a_*=0$ black hole has a similar shape compared with that of the spectra which cannot be well fit in the $a_*=0.5$ cases. Whether the corona temperature $kT_e$ is a free parameter or not, it does not significantly improve these fits, therefore abnormal trends of disk fraction and $\Gamma$ and the bad fits seen in lower spin cases are more likely caused by the geometrical effects of a limited-size corona, instead of the failure of modeling the cut-off energy.

\subsection{Impact of the coronal temperature}\label{temperature_discussion}

To study the influence of the coronal temperature on the output spectra, we simulate a $R_{\rm{out}}=30~R_{\rm{g}}$ sandwich corona with varying optical depth and different coronal temperatures (25~keV, 50~keV, 100~keV, and 300~keV). There are mainly two reasons for this choice: i) with the largest outer radius in our parameter space of the sandwich corona, the abundant scattered photons should manifest the impact of temperature; ii) as shown by Fig.~\ref{sandwich_grid}, the $R_{\rm{out}}=30~R_{\rm{g}}$ sandwich corona at 50~keV can be possible for both the soft and hard state changing from the minimum to maximum optical depth. We can use this part of the parameter space to test how the coronal temperature will change the boundary of different states.

\begin{figure*}
    \centering
    \includegraphics[width=0.4\linewidth]{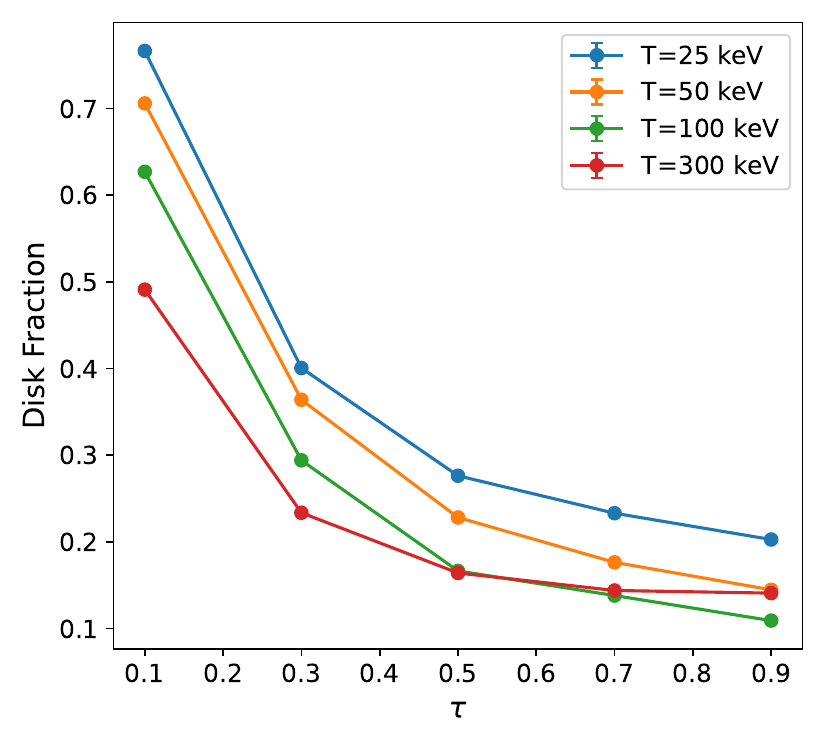}
    \includegraphics[width=0.4\linewidth]{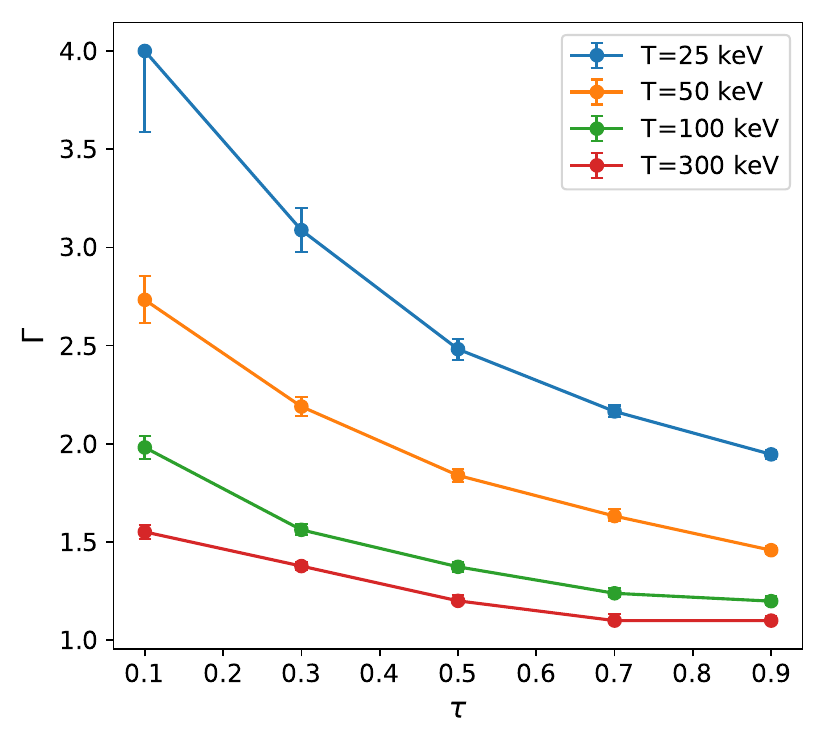}
    \caption{Change of the disk fraction (left panel) and photon index $\Gamma$ (right panel) with the optical depth $\tau$ for a $R_{\rm{out}}=30~R_{\rm{g}}$ sandwich corona with different temperature around a $a_*=0.998$ black hole. The disk fraction values for $T=300~keV$ and $\tau=0.7, 0.9$ are a bit higher than expected because for these points the fitting values of $\Gamma$ hit the lower boundary 1.1 and the fits are not so good.}
    \label{temperature}
\end{figure*}

The results are shown in Fig.~\ref{temperature}. If we increase the coronal temperature, we reduce the values of the disk fraction and $\Gamma$. This is because the energy a photon can be Compton-scattered to depends on the temperature of the electron scattering it. Therefore the temperature of the corona changes the slope of the power law and the fraction of the energy flux between the disk and corona emission. For our fixed $R_{\rm{out}}$, the corona with very low temperature ($\lesssim25$~keV) can never describe the hard state and the corona with very high temperature ($\gtrsim100$~keV) can never describe only the soft state\footnote{We note that a lamppost corona does not seem to be able to describe the hard state even if we change its temperature, because we cannot significantly decrease the disk fraction to a level suitable for the hard state ($< 25$\%).}.  

The changes in the coronal size and temperature are both important for the different spectral properties we see in different states, as has also been shown from observations. In this work we mainly focus on the influence of geometrical changes, but a larger simulation parameter space with changing corona size and temperature will definitely help us to get a more complete understanding, which may be the direction of future work.

\subsection{\textit{IXPE}: breaking the degeneracy of energy spectra}\label{IXPE}

We infer the possible parameter space of the hard state and soft state for different coronal geometries in Section~\ref{result}. However, there are some cases in which the disk fraction and photon index of a sandwich and spherical corona can be the same, making the energy spectra not distinguishable. The soft state of a lamppost corona is not discussed here because the significant dominance of the disk emission makes it well distinguishable from the other two geometries.

\begin{figure*}
    \centering
    \includegraphics[width=0.48\linewidth]{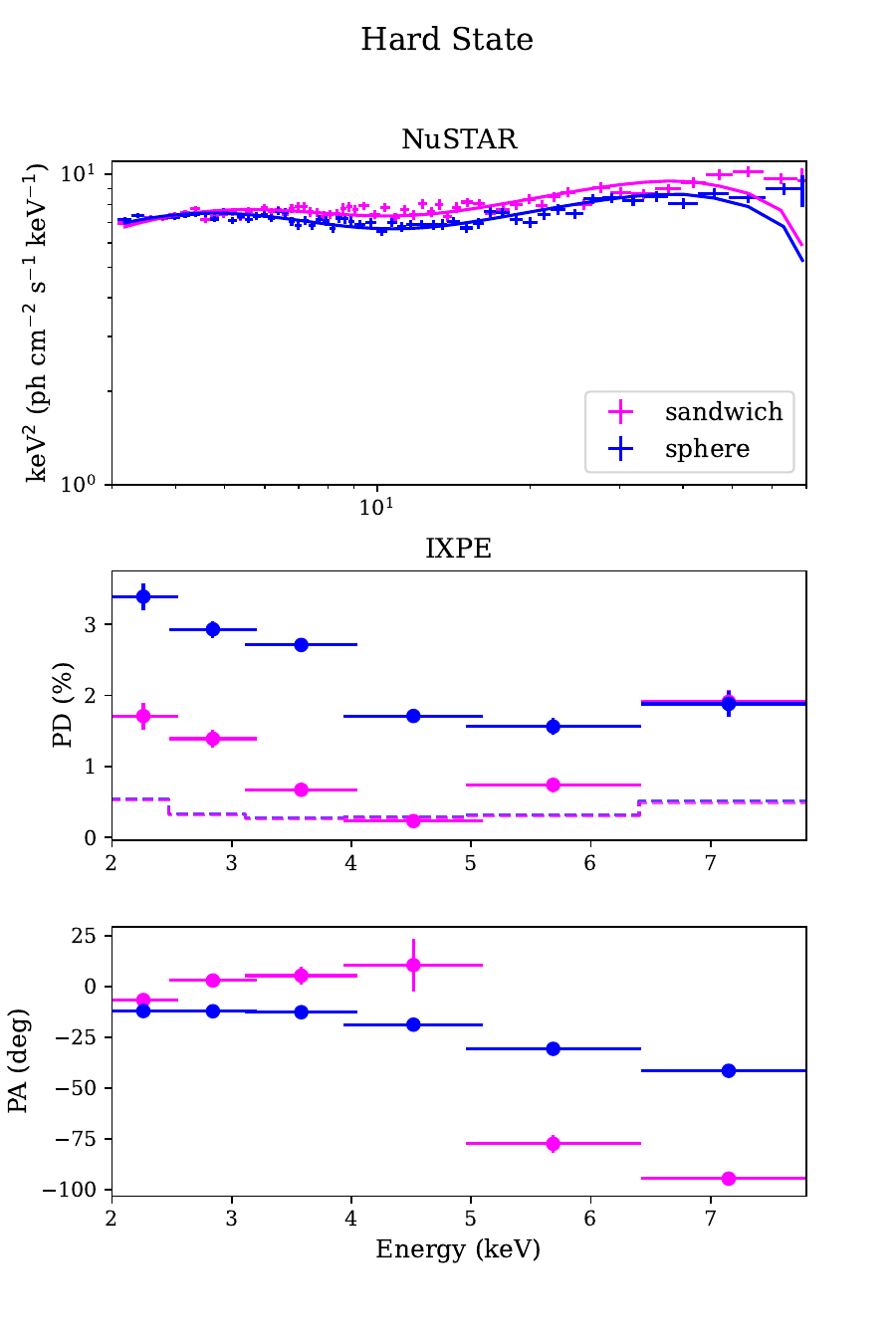}
    \includegraphics[width=0.48\linewidth]{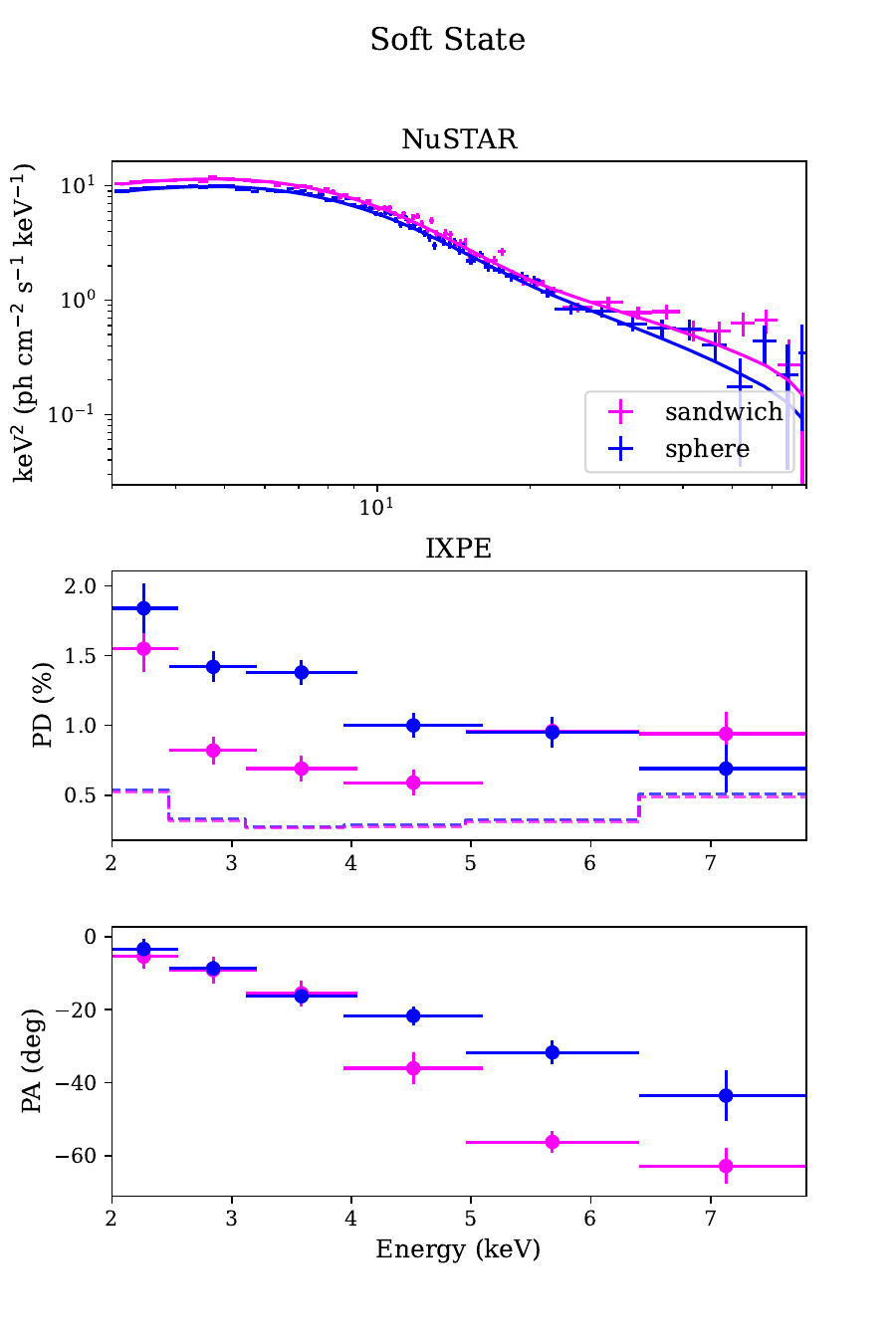}
    \caption{Left panels: the selected hard spectra of the sandwich and spherical corona corresponding to the lower triangles in Fig.~\ref{sandwich_grid} and Fig.~\ref{sphere_grid}. Right panels: the selected soft spectra of the sandwich and spherical corona corresponding to the upper triangles in Fig.~\ref{sandwich_grid} and Fig.~\ref{sphere_grid}. The upper panels show the simulated 30~ks \textit{NuSTAR} observations of the energy spectra with the error bars showing the simulated data, and the solid lines showing the best-fit \texttt{simplcut*kerrbb} model. The middle and lower panels show the simulated 100~ks \textit{IXPE} observations of the PD and PA spectra. The dashed lines in the middle panels present the $MDP_{99}$ in different energy bins. See the text for more details of the simulation settings.}
    \label{pol_compare}
\end{figure*}

We pick out a hard-state spectrum of the sandwich geometry and a hard-state spectrum of the spherical geometry with similar spectral parameters (noted by the lower triangles in Fig.~\ref{sandwich_grid} and Fig.~\ref{sphere_grid}) and plot them in the upper left panel of Fig.~\ref{pol_compare}. The soft-state spectra for different geometries (noted by the upper triangles in Fig.~\ref{sandwich_grid} and Fig.~\ref{sphere_grid}) are shown in the right upper panel of Fig.~\ref{pol_compare}. For the selected hard spectra, the spectrum of a $R_{\rm{out}}=30~R_{\rm{g}}$ and $\tau=0.6$ sandwich corona has disk fraction $\sim$ 22\% and $\Gamma=1.72\pm0.03$, while that of a $R=30~R_{\rm{g}}$ and $\tau=2.5$ spherical corona has disk fraction $\sim$ 25\% and $\Gamma=1.73\pm0.03$. For the selected soft spectra, the spectrum of a $R_{\rm{out}}=18~R_{\rm{g}}$ and $\tau=0.1$ sandwich corona has disk fraction $\sim$ 73\% and $\Gamma=3.0\pm0.2$, while that of a $R=12~R_{\rm{g}}$ and $\tau=0.5$ spherical corona has disk fraction $\sim$ 70\% and $\Gamma=3.2\pm0.2$.

Here, we want to check if the polarization observations of \textit{IXPE} can be an effective tool to break the degeneracy in the energy spectra. We put the simulated energy, PA and PD spectra with \texttt{MONK} into the \texttt{ixpeobssim} v31.1.0\footnote{\url{https://ixpeobssim.readthedocs.io/en/latest/index.html}} software as the point source model\footnote{\url{https://ixpeobssim.readthedocs.io/en/latest/source_models.html}}. Then we use the \texttt{ixpobssim}\footnote{\url{https://ixpeobssim.readthedocs.io/en/latest/pipeline.html}} tool to simulate 100~ks \textit{IXPE} observations of the source model. The resulting polarimetric spectra are shown in the middle and lower panels of Fig.~\ref{pol_compare}. While the selected energy spectra of different coronal geometries in different states strongly overlap with each other, the PD and PA spectra of a 100~ks \textit{IXPE} observation are distinguishable, and the PD values are well above the minimum detectable polarization at 99\% confidence level ($MDP_{99}$) in most energy bins. We calculate the likelihood ratio between the photon angular distributions derived from the PD and PA spectra of different coronal geometries, and find the two observations are quite distinct at $>5\sigma$ level. This shows the potential of \textit{IXPE} to break the degeneracy in the energy spectra and constrain the exact coronal geometry in a certain state.

\section{conclusion}\label{conclusion}

This study aims to understand the possible corona scenario in the hard state and soft state of BHXRBs. We use the Monte Carlo code \texttt{MONK} to simulate the spectra of three coronal geometries: the sandwich, spherical and lamppost corona, and fit the spectra with \textit{NuSTAR} response using the model \texttt{simplcut*kerrbb}. From the photon index $\Gamma$ and the disk fraction, we infer the possible parameter space of different geometries in different spectral states. The main conclusions are as follows:

\begin{itemize}
    \item In the sandwich geometry, the disk fraction decreases as $\tau$ and $R_{\rm{out}}$ increase, and $\Gamma$ decreases with $\tau$. The soft state occurs when the optical depth $\lesssim$ 0.15, while the hard state occurs when the optical depth is $\sim$ 0.5--0.9 and the corona outer radius $\gtrsim18~R_{\rm{g}}$. The hieght and thickness of the corona do not have a significant impact on the results.
    \item In the spherical geometry, the changing trend of disk fraction and $\Gamma$ with $\tau$ and $R$ is similar to that in the sandwich geometry. The soft state occurs in the lower left part of the parameter space bounded by $\tau=1.5, R=6~R_{\rm{g}}$ and $\tau=0.5, R=12~R_{\rm{g}}$, and the line connecting those two points. In the hard state, the spherical corona should have $\tau\sim$ 2--3 and $R\gtrsim18~R_{\rm{g}}$.
    \item In the lamppost geometry, the disk emission is always dominant in the simulation parameter space. The insufficient scattering of disk photons makes a lamppost corona (with a size comparable to or smaller than $\sim5~R{\rm{g}}$) impossible for the hard state. $\Gamma$ cannot be well constrained when the the optical depth is low $\tau \lesssim2$ or the height is high $h\gtrsim3~R_g$. While a harder spectrum may be possible for extended or hybrid lamppost-like geometries, we can certainly rule out very compact lamppost coronae.
    \item The trend of $\Gamma$ from \texttt{MONK} simulation and fitting results of the sandwich and spherical geometry are found to be generally similar to the \texttt{CompTT} model. However, the values of $\Gamma$ deviate when the optical depth is relatively high ($
    \sim0.5$ for the sandwich geometry and $\sim2.0$ for the spherical geometry).
    \item When the spectral properties are close in the sandwich and spherical geometry, the \textit{IXPE} observations of PA and PD have the potential to further distinguish the exact the coronal geometry in one certain state.
\end{itemize}

We further discuss the impact of the inclination angle, black hole spin and coronal temperature in the sandwich geometry, and find that:

\begin{itemize}
    \item For different inclination angles, the changing trend of $\Gamma$ and disk fraction with $\tau$ and $R_{\rm{out}}$ is similar, but their absolute values differ because the effective optical depths and covering fractions for photons observed at different inclination angles differ. 
    \item For a black hole with lower spin ($a_*=0.5$ or $0$), the ISCO is larger, therefore the corona with the same outer radius covers a smaller disk fraction compared with the maximum spinning black hole. As a result, we observe strong geometrical effects and the spectra cannot be fit well when $\tau$ is high and $R_{\rm{out}}$ is not large enough. The residuals of the fits show that the disk emission is distorted and the power-law tail is underestimated in these cases.
    \item When the coronal temperature increases, the disk fraction and $\Gamma$ decreases, showing that both the coronal size and temperature play a role in the exact shape of a spectrum in different states. Future work with simulations over a larger parameter space, where the coronal geometrical properties and the temperature can change simultaneously, will help us gain a more complete understanding of this problem.
    
\end{itemize}

\vspace{0.5cm}

{\bf Acknowledgments --}
We thank the anonymous referee for constructive comments and suggestions, which have significantly improved the quality of this manuscript. The work of N.F. and C.B. was supported by the National Natural Science Foundation of China (NSFC), Grant No.~12250610185, W2531002, and 12261131497. N.F. acknowledges the support by CURE (Hui-Chun Chin and Tsung-Dao Lee Chinese Undergraduate Research Endowment) (24921), and National Undergraduate Training Program on Innovation and Entrepteneurship grant No.~202410246141.

\bibliographystyle{aasjournal}
\bibliography{ref}

\end{document}